\documentclass[aps,prapplied,reprint,superscriptaddress,longbibliography]{revtex4-1}

\usepackage{amsmath}
\usepackage{graphicx}
\usepackage{gensymb}
\usepackage{textcomp}

\begin{document}


\title{Thermal-error regime in high-accuracy gigahertz single-electron pumping}



\author{R.~Zhao}
\email[]{ruichen.zhao@student.unsw.edu.au}
\affiliation{School of Electrical Engineering and Telecommunications, University of New South Wales, Sydney, New South Wales 2052, Australia}

\author{A.~Rossi}
\affiliation{Cavendish Laboratory, University of Cambridge, J J Thomson Avenue, Cambridge CB3 0HE, United Kingdom}

\author{S.~P.~Giblin}
\affiliation{National Physical Laboratory, Hampton Road, Teddington, Middlesex TW11 0LW, United Kingdom}

\author{J.~D.~Fletcher}
\affiliation{National Physical Laboratory, Hampton Road, Teddington, Middlesex TW11 0LW, United Kingdom}

\author{F.~E.~Hudson}
\affiliation{School of Electrical Engineering and Telecommunications, University of New South Wales, Sydney, New South Wales 2052, Australia}

\author{M.~M\"ott\"onen}
\affiliation{QCD Labs, COMP Centre of Excellence, Department of Applied Physics, Aalto University, 00076 AALTO, Finland}

\author{M.~Kataoka}
\affiliation{National Physical Laboratory, Hampton Road, Teddington, Middlesex TW11 0LW, United Kingdom}

\author{A.~S.~Dzurak}
\affiliation{School of Electrical Engineering and Telecommunications, University of New South Wales, Sydney, New South Wales 2052, Australia}


\date{\today}

\begin{abstract}

Single-electron pumps based on semiconductor quantum dots are promising candidates for the emerging quantum standard of electrical current. They can transfer discrete charges with part-per-million (ppm) precision in nanosecond time scales. Here, we employ a metal-oxide-semiconductor silicon quantum dot to experimentally demonstrate high-accuracy gigahertz single-electron pumping in the regime where the number of electrons trapped in the dot is determined by the thermal distribution in the reservoir leads. In a measurement with traceability to primary voltage and resistance standards, the averaged pump current over the quantized plateau, driven by a \mbox{$1$-GHz} sinusoidal wave in the absence of magnetic field, is equal to the ideal value of $ef$ within a measurement uncertainty as low as $0.27$~ppm.

   
\end{abstract}


\maketitle

\section{Introduction}
Single-electron (SE) pumps can generate quantized electrical current by controlling the transport of individual electrons with an external periodic drive \cite{PekolaRevModPhys,SlavaProgPhys}. These devices relate the pumped direct current, $I$, to the elementary charge, $e$, and the driving frequency, $f$, through the expression, $I = nef$, where $n$ is an integer.  As an on-demand SE source, they can be useful in the context of quantum information processing as well as in the study of fermionic optics \cite{TimeofFlight,ClockedEmission,ubbelohde2015partitioning}. Arguably, the most important application of this technology is to realize a quantum standard of electrical current \cite{Metrologia2000}.

Single-electron pumps and turnstiles have been realized in various physical systems, including normal-metal tunnel junction devices \cite{keller1996accuracy,keller1999capacitance}, surface accoustic wave devices \cite{shilton1996high,talyanskii1997single}, superconducting devices \cite{niskanen2005evidence,vartiainen2007nanoampere,mottonen2008experimental}, hybrid superconductor-normal-metal turnstiles \cite{pekola2008hybrid}, quantum dots \cite {blumenthal2007gigahertz,jehl2013hybrid,NPL2012,PTB2015,giblin2016high,connolly2013gigahertz,fujiwara2004current,fujiwara2008nanoampere,ono2003electron,chan2011single,Ale2014,tanttu2016three,NTT2016} and single dopants or traps \cite{yamahata2017high,yamahata2014gigahertz,tettamanzi2014charge,lansbergen2012donor,roche2013two}. The tunable-barrier SE pumps based on semiconductor quantum dots (QDs) stand out from the competing technologies for providing a good balance between low pumping error and high output current \cite{NPL2012,Ale2014,NTT2016,KRISS2015,stein2016robustness}. 

Three different designs of GaAs pumps have achieved relative errors close to or below 1 part per million (ppm) in high-accuracy measurements traceable to primary standards \cite{NPL2012,KRISS2015,PTB2015,stein2016robustness,giblin2017robust}.
 These GaAs pumps transport a fixed number of electrons per cycle following a series of sequential back-tunneling events, known as the decay cascade \cite{decaycascade}. Previous studies indicate that strong magnetic fields, tailored waveform drives, and subkelvin temperatures are required for the GaAs pumps to achieve ppm level accuracy at gigahertz pumping frequencies \cite{NPL2012,KRISS2015,stein2016robustness}. These requirements render the realization of the quantum current standard demanding and restrict the user base of GaAs pump technology.

In contrast, QD pumps in silicon alleviate some of these burdens. Compared to depletion mode GaAs QDs, the gate-voltage-induced silicon QDs tend to have a larger addition energy due to their smaller physical size. This feature of the compact silicon devices enables accurate high frequency SE pumping in the decay-cascade regime without arbitrary waveform drives or high magnetic fields \cite{NTT2016,Ale2014}. The remarkable results recently achieved in silicon devices not only demonstrate the universality of SE pumping in tunable-barrier QDs at sub-ppm uncertainty, but also clearly indicate that a compact silicon SE pump may pave the way towards a more practical quantum standard of electrical current \cite{NTT2016}. From a pragmatic point of view, it is advantageous to implement the quantum current standard in silicon, since it is compatible with the metal-oxide-semiconductor (MOS) technology widely employed in industry. Through well-established fabrication techniques, silicon SE pumps can be seamlessly integrated with peripheral control circuits to deliver a cost-effective on-chip current standard.

One challenge for the SE pumps is that the large rf drive amplitude usually required at gigahertz pumping frequency may heat the electron reservoir up to several kelvins and result in excessive thermal errors \cite{GentoCrossOver,chan2011single}. When the electron reservoir temperature increases, forward tunneling of thermally excited electrons from the reservoir into the QD becomes significant during the charge capturing process, and the number of electrons trapped in the QD reflects the Fermi distribution of electrons in the leads \cite{kashcheyevs2014modeling,lukasCountingStat}. To the best of our knowledge, gigahertz high-accuracy SE pumping in the thermal regime has not been achieved among the silicon devices.

In this work, we use a silicon QD, fabricated employing a MOS planar gate stack technology \cite{angus2007gate,rossi2015silicon}, to demonstrate high-accuracy SE pumping in the regime where the number of pumped electrons in each cycle is determined by the thermal distribution of electrons in the reservoir leads. We investigate whether the accuracy of our SE pump significantly deteriorates due to drive-induced heating in the electron reservoir, as reported in previous studies \cite{chan2011single,GentoCrossOver}. Fits of the measurement data to the thermal model of electron capture yield a theoretical lower bound of 4 parts per billion (ppb)
 for the thermal error on the $ef$ current plateau at $f=1$~GHz. In addition, we experimentally measure the pumped current using a high-accuracy measurement set-up, which compares the pumped current to a reference current derived from primary voltage and resistance standards \cite{NPL2012}. We find that the averaged current on the plateau, induced by a sine wave drive in the absence of magnetic field, matches the $ef$ value within the measurement uncertainty of $\sim 0.3$~ppm. This is the most accurate measurement of the current from a silicon electron pump to date.

\section{Device Architecture and Experimental Methods}

\begin{figure}

\includegraphics[width=0.7\linewidth]{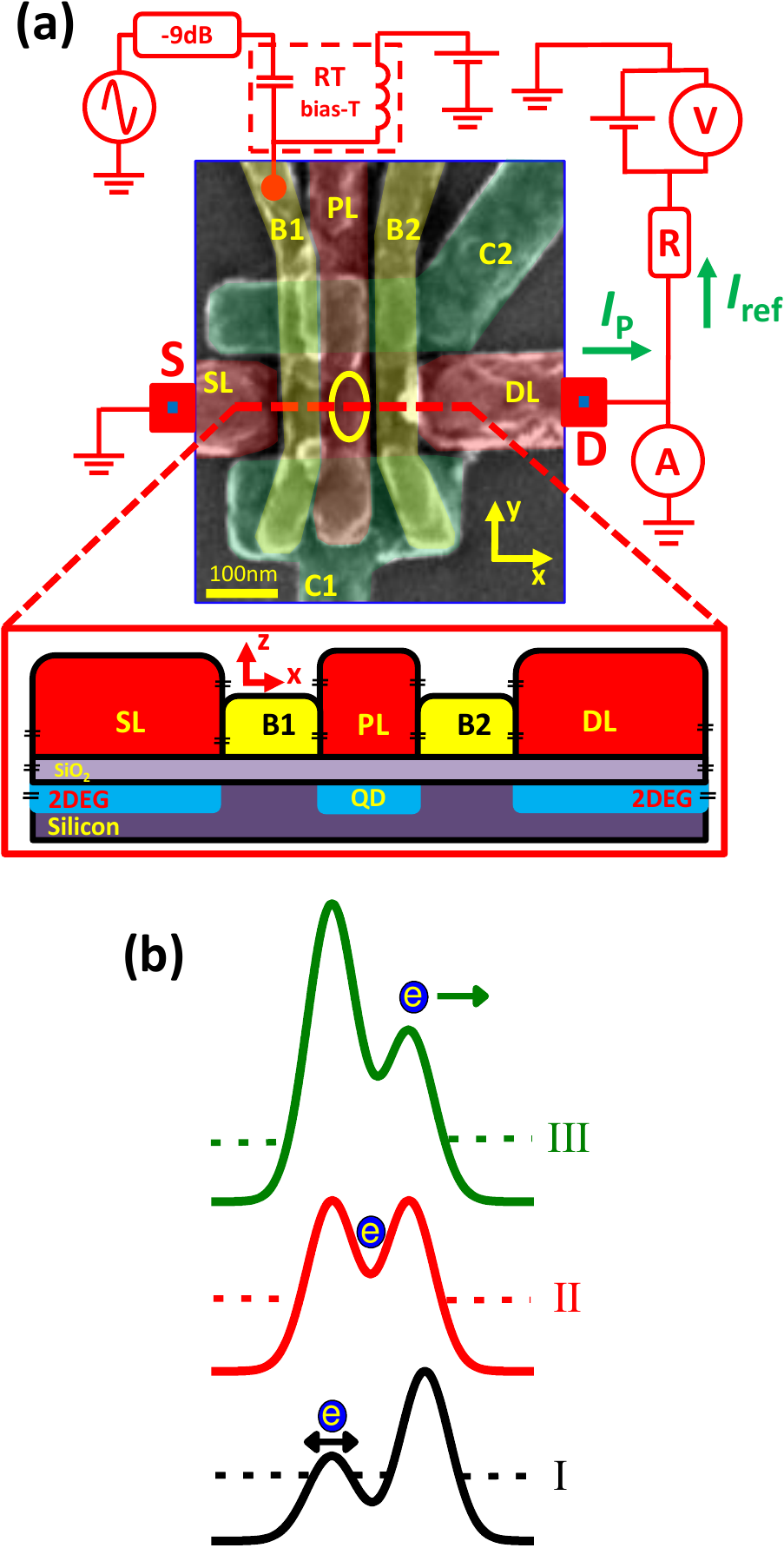}%
\caption{(a) False-color scanning electron microscope image of an electron pump similar to the one used. The yellow circle highlights the approximate region where a quantum dot is formed. A schematic of the measurement set-up as well as an illustrative cross-sectional view of the metal-oxide-semiconductor structure are also shown. The drain contact is connected to the reference current source used in the high-accuracy measurements. It consists of a temperature controlled 1-G$\Omega$ thick-film resistor and a voltage source (Keithley213) monitored by a high-accuracy voltmeter (HP 3458A). (b) Sketch of the conduction band energy profile (solid lines) and fermi-level (dashed lines) during a pumping cycle.\label{fig:DeviceAndSetup}}

\end{figure}

\begin{figure}
	
	\includegraphics[width=0.8\linewidth]{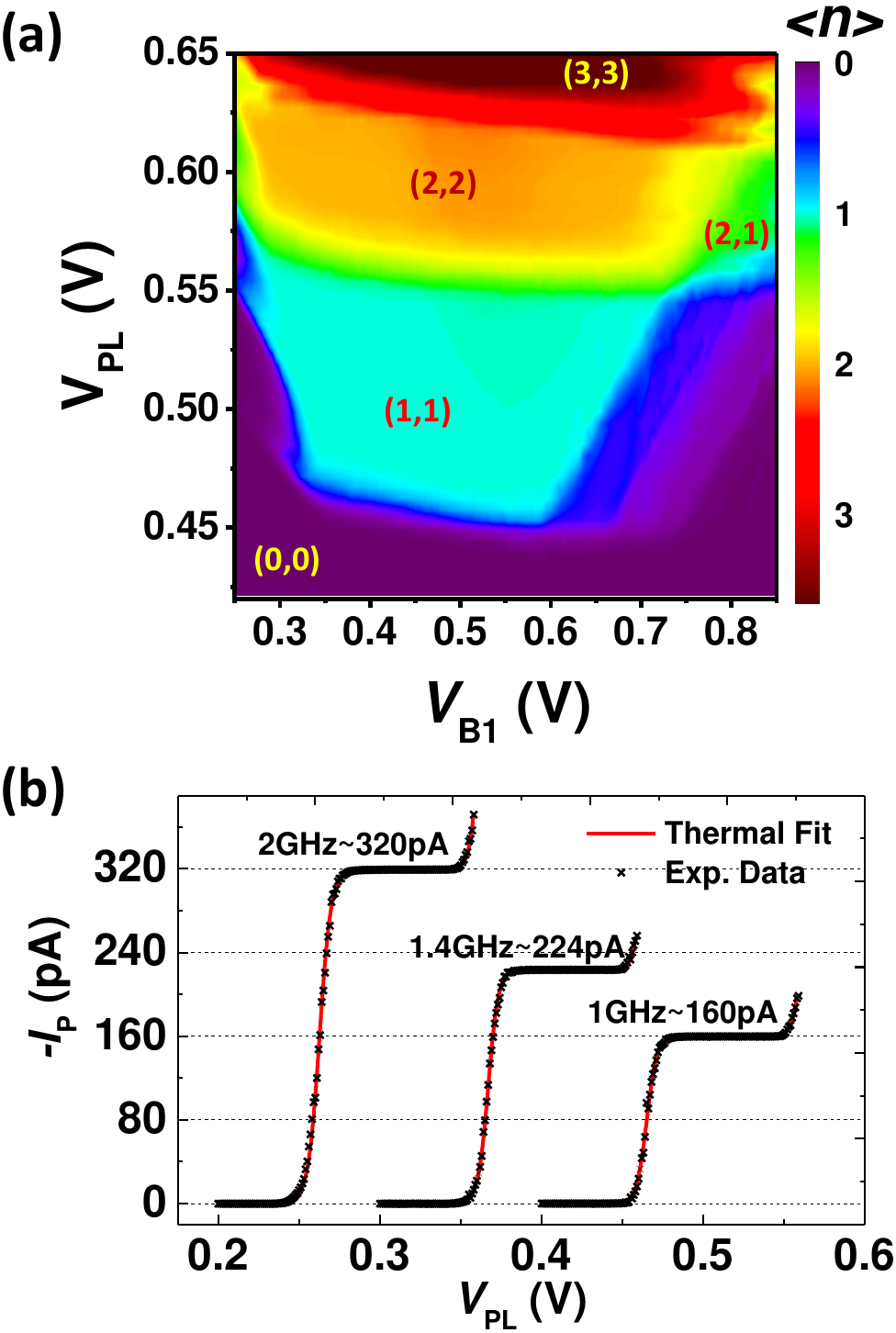}%
	\caption{(a) Coarsely tuned current plateaux at $f=500$~MHz measured using the normal-accuracy set-up. In the notation ($m$,$n$), $m$ ($n$) represents the ideal number of captured (ejected) electrons. Here, $V_{\textmd{SL}} = V_{\textmd{B2}} = 1.5$ V, $V_{\textmd{DL}} = 1.75$ V, $V_{\textmd{C1}} = -1$ V, $V_{\textmd{C2}} = 0.2$~V, and $P_{\textmd{B1}}=2$ dBm. (b) Normal-accuracy measurements (black crosses) of pumped current as a function of the plunger gate voltage for different pumping frequencies and fits to the thermal model (red solid lines). Data have been horizontally shifted for clarity. Parameter settings: $V_{\textmd{SL}} = V_{\textmd{B2}} = 1.5$ V, $V_{\textmd{B1}} = 0.45$ V, $V_{\textmd{DL}} = 1.9$ V, $V_{\textmd{C1}} = -1.04$ V, $V_{\textmd{C2}} = 0.187$ V and $ P_{\textmd{B1}}=3$ dBm. \label{fig:PumpMap}}
	
\end{figure}

The sample used in the experiments was fabricated on a high-purity near-intrinsic silicon wafer. We thermally grow 7-nm high-quality SiO$_2$ gate oxide on top of the substrate. Three layers of aluminium gate electrodes are lithographically defined on top of the gate oxide. Between each layer, the sample is heated up to $150\,^{\circ}\mathrm{C}$ in air to form an aluminium oxide coating on the electrode surface. This coating provides good electrical insulation between different metal layers \cite{rossi2015silicon,angus2007gate}.

A scanning electron microscope image of the aluminum gate stack of a device similar to the one used in the experiments is shown in Fig.~\ref{fig:DeviceAndSetup}(a). These metal gates, connected to programmable dc voltage sources through \mbox{$200$-Hz} low-pass filters, can locally induce two dimensional electron gas (2DEG) channels or potential barriers at the Si/SiO$_2$ interface. By tuning the individual gate voltages, a quantum dot containing a few conduction electrons can be defined below the plunger gate (PL) as shown in Fig.~\ref{fig:DeviceAndSetup}(a). Electron reservoirs are accumulated below the source lead (SL) and the drain lead (DL), electrically connecting the quantum dot to the ohmic contacts. 

We optimized the pump performance using a normal-accuracy measurement set-up shown in Fig.~\ref{fig:DeviceAndSetup}(a). The pumped current, $I_\textmd{P}$, is measured by a low-noise transimpedance amplifier (Femto DDPCA300) connected to the drain contact. The reference current source used in high-accuracy measurement is also connected to the drain, but it is switched off ($V=0$) in the normal-accuracy measurement set-up. We operate the SE pump with a sinusoidal excitation. As shown in Fig.~\ref{fig:DeviceAndSetup}(b), each pumping cycle begins with the rf drive lowering the potential barrier between the QD and the source reservoir and loading the QD with electrons. Then the rf drive raises the barrier to trap electrons and eject some or all of them to the drain reservoir. Gate B1 was driven by a microwave source (HP8341B) through a room temperature bias-tee followed by a \mbox{9-dB} attenuator. The source was synchronized to a 10-MHz reference frequency derived from a primary caesium frequency standard. All RF power levels quoted in this paper refer to the power after the \mbox{$9$-dB} attenuator. All measurements presented in this work were carried out on a single device in the absence of a magnetic field with a small ($\sim 250$ $\mu$V) stray bias across the pump due to the current preamplifier. The sample was cooled in a helium-3 cryostat with a base temperature of $300$~mK.

We take the following approach to search for a stable low-error current plateau: First, the capacitive coupling strength of the quantum dot to each gate is obtained from the period of the corresponding Coulomb blockade oscillations. Second, the two gate voltages that have the strongest capacitive coupling to the dot potential, namely, $V_{\textmd{B1}}$ and $V_{\textmd{PL}}$, are selected to be the main sweep parameters. Third, a sinusoidal excitation with a relatively low frequency, starting from \mbox{$500$~MHz}, is applied to B1. We gradually increase the rf drive power, $P_\textmd{B1}$, until a plateau structure, shown in Fig.~\ref{fig:PumpMap}(a), appears in the $V_{\textmd{B1}}$ -- $V_{\textmd{PL}}$ plane. Finally, we decrease $V_\textmd{C1}$ and $V_\textmd{C2}$ to obtain a flatter current plateau \cite{Ale2014}. We verify the robustness of the well optimized current plateau at high pumping frequencies. As shown in Fig.~\ref{fig:PumpMap}(b), the $ef$ current plateau is well pronounced up to \mbox{$2$~GHz} without changing the gate voltages or rf power. 

The search time is determined by the scan speed of the normal-accuracy measurement set-up, which is limited by the \mbox{$200$-Hz} low-pass filters connected between the dc voltage sources and the metal gates in this study. The tune-up process lasted a few hours and was performed only once during the whole measurement campaign. The fine-tuned current plateau, presented in Fig.~\ref{fig:PumpMap}(b), was stable throughout the high-accuracy measurement over a time period of a few weeks. Using this tune-up procedure, tens of devices with identical design have showed high frequency current plateaux. Although these devices showed extremely low theoretical error rates and excellent stability over time, due to the limited access to the high-accuracy measurement setup, this latest study is the only one where we could experimentally determine the pumping accuracy at the sub-ppm level.

\begin{figure}
	
	\includegraphics[width=0.73\linewidth]{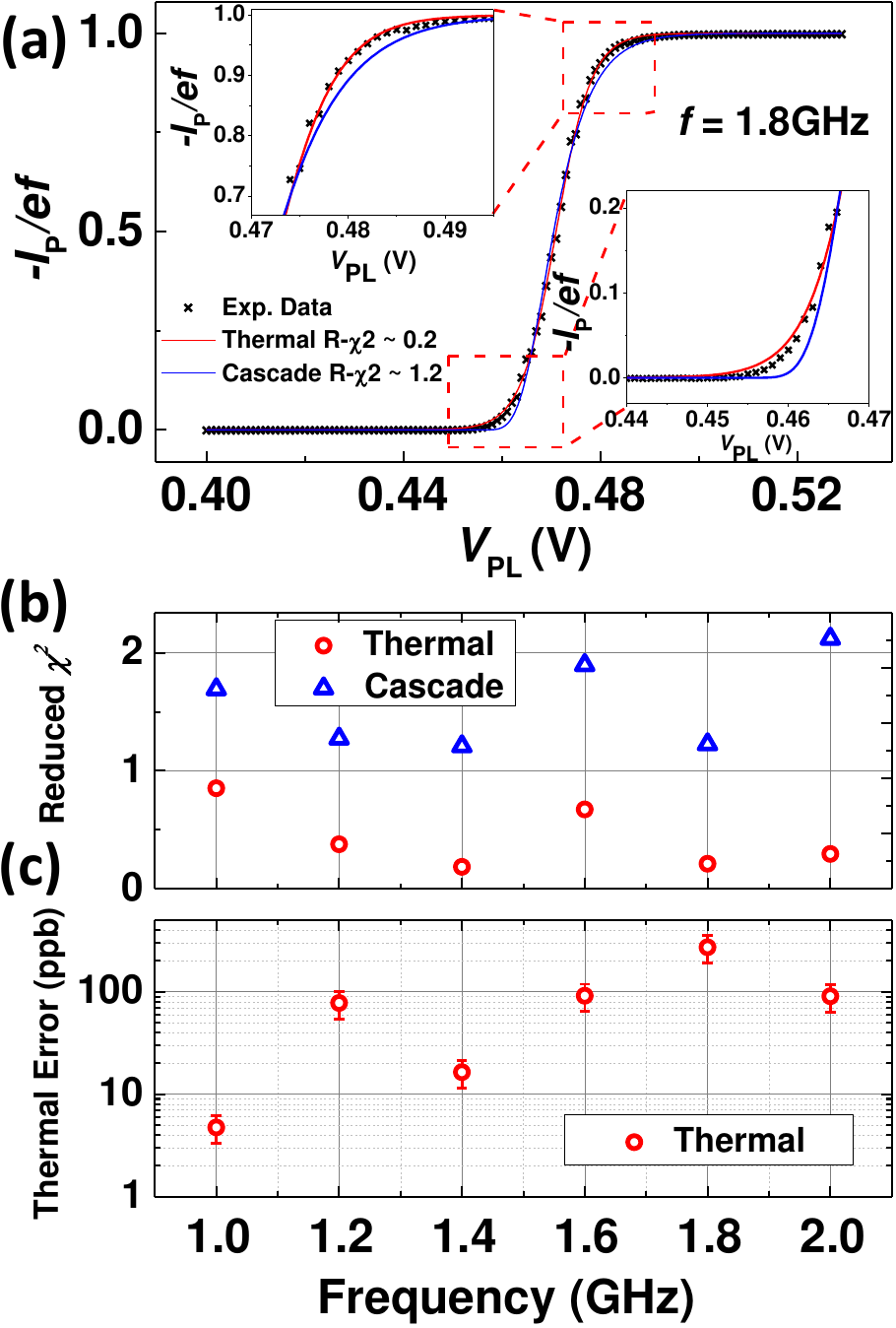}%
	\caption{(a) Normal-accuracy measurement (black crosses) of pumped current as a function of plunger gate voltage and its fits to the thermal (red solid line) and decay-cascade models (blue solid line). Insets: selected data from the main panel on expanded axes. (b) The reduced $\chi ^2$ fit error as a function of pumping frequency for both thermal and decay-cascade models. (c) The thermal error at the center of the first plateau predicted according to the fit to the thermal model at different frequencies. The plateau center is defined as the point of inflection of the thermal fit. The red error bar represents the typical error of the fit ($\sim$10\%). For (a), (b), and (c), all gate voltage settings and rf power level are identical to Fig.~\ref{fig:PumpMap}(b). \label{fig:Plateaux}}
	
\end{figure}

\section{Results}

\subsection{SE Pumping in the Thermal Regime}

The shape of the current staircase between two adjacent plateaux as a function of the QD depth-tuning gate $V_\textmd{PL}$ (Fig.~\ref{fig:PumpMap}(b)) provides information about the process by which the QD is decoupled from the source lead. To date, the reported accurate semiconductor pumps \cite{PTB2015,NTT2016,NPL2012,KRISS2015,giblin2017robust} have operated in the decay-cascade regime \cite{decaycascade}, where the final
number of electrons in the QD is determined by a
one-way cascade of back-tunneling events \cite{decaycascade}. Consequently, the average number of captured electrons, $\langle m \rangle$, is characterized experimentally by an asymmetric staircase modelled using a double-exponential function of the QD depth-tuning gate voltage, which in our device is $V_\textmd{PL}$,
\begin{equation}
\begin{split}
\langle m \rangle =\sum_{n} \exp(-\exp(-aV_{\textmd{PL}}+\Delta_\textmd{n})),
\end{split}
\label{eq:decaycascade}
\end{equation} 
where $a$ and $\Delta_\textmd{n}$ are fit parameters. 

In this work, we consider the possibility that the electron reservoir is heated by the large-amplitude sinusoidal drive, leading to charge capture in the thermal regime. In this regime, electrons are exchanged between the dot and the leads during the initialisation, so that the average number of captured electrons, $\langle m \rangle$, follows the grand canonical distribution \cite{GentoCrossOver} and can be expressed as 
\begin{equation}
\begin{split}
\langle m \rangle =\sum_{n} 1/\{1+\textmd{exp}[E_\textmd{add}^\textmd{(n)}/(k_\textmd{B}T)]\},
\end{split}
\label{eq:grandCanonical}
\end{equation}
where $k_\textmd{B}$ is the Boltzmann constant, $T$ is the electron temperature of the source resevoir, and $E_\textmd{add}^\textmd{(n)}$ is the addition energy of the $n$th electron. We assume that the addition energy is approximately a linear function of $V_\textmd{PL}$ within the small voltage range swept for pumping in the single electron regime. Therefore, Eq.~(\ref{eq:grandCanonical}) can be further expressed as a function of $V_\textmd{PL}$ 
\begin{equation}
\begin{split}
\langle m \rangle = \sum_{n} 1/[1+\textmd{exp}(A_\textmd{n}+B_\textmd{n} V_{\textmd{PL}})],
\end{split}
\label{eq:cascadedFermi}
\end{equation}
where $A_\textmd{n}$ and $B_\textmd{n}$ are the fit parameters for the $n$th current plateau. Assuming the ejection error is negligible during pumping, the normalized current, $-I_\textmd{P}/ef$, measures the average number of captured electrons. In this work, the normalized pumped current, $-I_\textmd{P}/ef$, is used in the numerical fit of $\langle m \rangle$ for both decay-cascade and thermal models.

As shown in Fig.~\ref{fig:Plateaux}(a), the current staircase of our device is more accurately described by the thermal model than the decay-cascade model. The reduced $\chi^2$ fit error for the thermal model, displayed in Fig.~\ref{fig:Plateaux}(b), is significantly lower than that for the decay-cascade model for all studied pumping frequencies. This strongly suggests our device is indeed operating in the thermal regime.

We estimate the reservoir electron temperature for the measurement in Fig.~\ref{fig:Plateaux} using the following method. We extract the ratio of $E_\textmd{add}/(k_\textmd{B}T)$ from the thermal fits presented in Fig.~\ref{fig:Plateaux}. Along with an addition energy of $17$~meV, calculated based on a conduction band profile simulated in the commercial semiconductor software package ISE-TCAD\cite{tcad}, we deduce the local electron temperature near the SE pump to be around $9$~K at a $f=1$~GHz. We need to estimate the addition energy using a simulation because the tunnel barriers are made completely opaque in the SE pumping regime in order to prevent co-tunneling errors \cite{pothier1992single}, which prevents the direct observation of $E_\textmd{add}$ in conductance measurements. More details on the TCAD simulation and the estimation of the QD addition energy are presented in the supplementary information \cite{suppinfo}. 

In our previous work \cite{Ale2014}, a device with similar design driven by a much smaller rf signal, rougly $-6$ dBm, demonstrated SE pumping in the decay-cascade regime. This suggests that the thermal regime observed in the present experiments is indeed due to heating of the electron reservoirs by the large rf drive signal. A similar heating effect has been observed in a SE shuttle fabricated employing the same silicon technology \cite{chan2011single}. An effective electron temperature of $7$~K, attributed to rf-induced heating, has also been reported in another SE pumping study employing a silicon nanowire device \cite{GentoCrossOver}.

Next, we investigate whether the accuracy of our SE pump will, as reported in previous studies using silicon devices \cite{GentoCrossOver,chan2011single}, significantly deteriorate due to such severe localized heating in the electron reservoir. Since our pump operated in the thermal regime, the main cause of capture error is expected to be thermal fluctuations of the QD electron number during its decoupling from the source reservoir \cite{GentoCrossOver}. The thermal error rate at the optimal working point of the $I=ef$ plateau can be estimated as $P_\textmd{error}^\textmd{thermal} = 1-1/\{1+\textmd{exp}[E_\textmd{add}^\textmd{(1)}/(k_\textmd{B}T)]\}$ \cite{GentoCrossOver}, with the optimal working point given by the point of inflection of the fit line. Figure \ref{fig:Plateaux}(c) shows the thermal error rate as a function of frequency. Despite the elevated electron temperature, we find the thermal error is as low as $4$~ppb for $f=1$~GHz. However, this should only be considered a lower bound for the overall error rate. Other error mechanisms, such as the non-adiabatic excitation of the captured electron \cite{Masaya2011PRL,Fletcher2012PRB,NTT2016,yamahata2017high} may be present and are not considered in the above analysis.

\begin{figure*}
	\includegraphics[width=\textwidth]{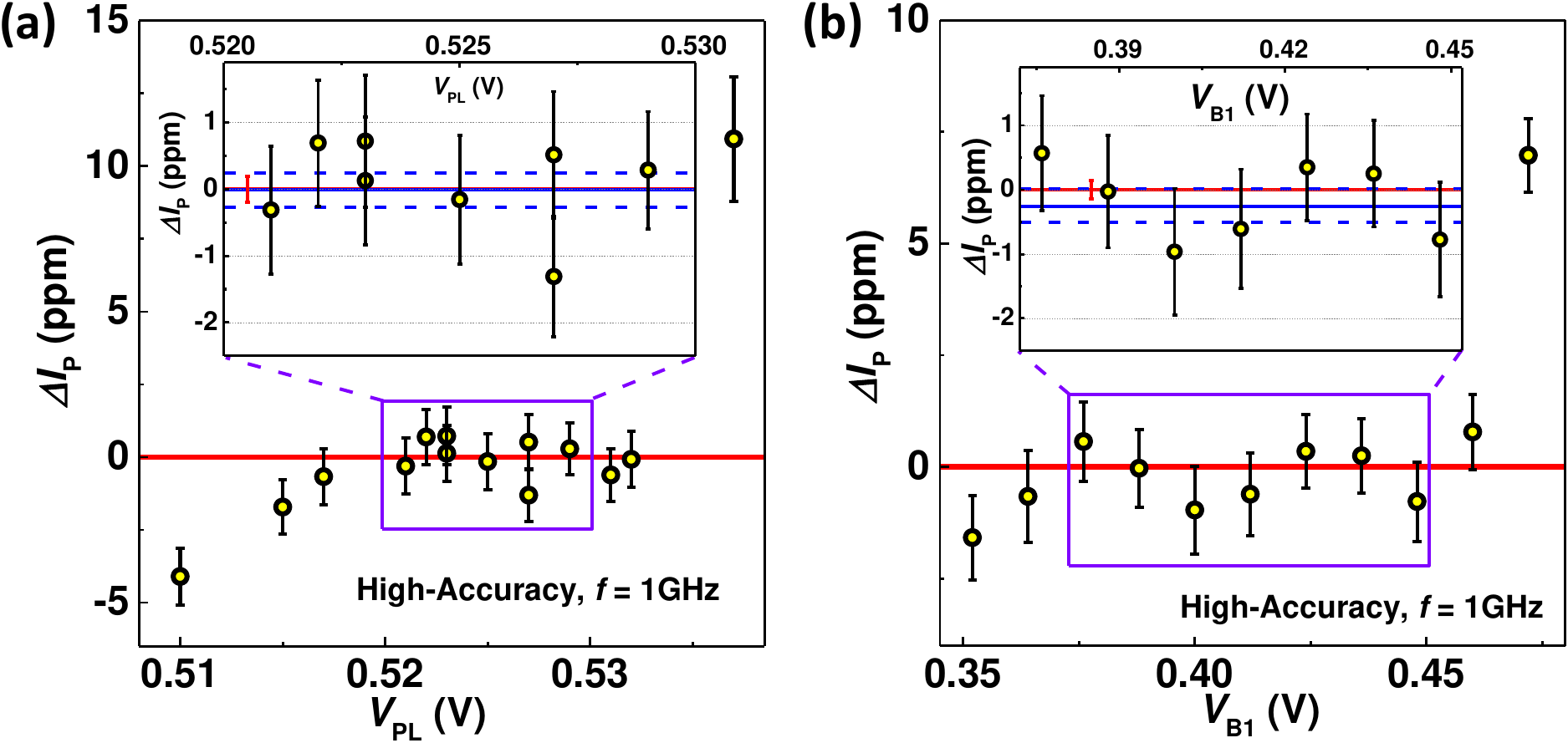}%
	\caption{ High-accuracy measurements of the relative deviation $\Delta I_{\textmd{P}}$ of the pumped current $I_{\textmd{P}}$ from the ideal $ef$ value as functions of (a) $V_{\textmd{PL}}$ and (b) $V_{\textmd{B1}}$ at $f=1$~GHz. Each data point is obtained by averaging $85$ minutes of raw data. The black error bar represents the 1$\sigma$ relative statistical uncertainty $U_{\text{ST}}$ over each $85$-min measurement, which is typically $\sim 0.9$~ppm. The red solid line indicates $\Delta I_\textmd{P}=0$. We define the current plateau as the region where the fit to the thermal model deviates from $ef$ by less than $\pm 0.03$~ppm. Insets: Data points in the current plateau region on expanded axes. The red error bar corresponds to the relative systematic uncertainty for the measurement of $I_\textmd{ref}$. The blue solid (dashed) line corresponds to the mean (error of the mean) of the points on the plateau. All the uncertainties quoted in this paper are $1\sigma$ and have been rounded up to the next $0.01$~ppm. The parameter settings are the same as in Fig.~\ref{fig:PumpMap}(b), except  $V_{\textmd{PL}} = 0.525$ V for (b).\label{fig:HighResMeas}}
\end{figure*}

\subsection{High-Accuracy Measurement}
To experimentally investigate the quantised current accuracy, we employ the high-accuracy measurement scheme described in Ref. \cite{NPL2012}. We compare the pumped current, $I_{\textmd{P}}$, to a reference current, $I_{\textmd{ref}}$, with traceability to primary voltage and resistance standards. The transimpedance amplifier is used to measure the difference between these currents, $I_\textmd{null}$. Because it measures a very small signal, the drift in gain of the transimpedance amplifier, for example due to temperature fluctuations, introduces only a small contribution to the overall uncertainty.

In previous studies employing the same measurement set-up \cite{NPL2012,KRISS2015,NTT2016}, the $0.8$~ppm systematic uncertainty in the calibration of the $1$~G$\Omega$ resistor was by far the dominant contribution to the uncertainty budget. In this work, we introduce a revised uncertainty budget following a first-principles re-evaluation of the cryogenic current comparator (CCC) bridge used to calibrate the resistor \cite{Giblin201xCCC}. In the revised uncertainty budget, the largest systematic term is $0.1$~ppm, due to the $10$~M$\Omega$ reference resistor used in the calibration, and the statistical uncertainty in the resistor calibration is also of order $0.1$~ppm. A recent comparison of precision reference current sources \cite{drung2015validation} has highlighted problems with short-term drift affecting high-value standard resistors. To reduce the impact of this drift on the pump measurements to well below $0.1$~ppm, in this work we calibrated the $1$~G$\Omega$ resistor very frequently, with an interval between calibrations as short as $2$~days.

We carried out our high-accuracy measurement on an optimized $I=ef$ plateau at 1 GHz. The pumped current as a function of $V_\textmd{PL}$ is shown in Fig.~\ref{fig:HighResMeas}(a), where the fractional deviation of the pumped current from $ef$ is defined as  $\Delta I_{\text{P}} \equiv (I_{\text{P}}-e_{\text{90}}f)/e_{\text{90}}f$. We use $e_{\text{90}} \equiv 2/(R_{\text{K-90}}K_{\text{J-90}})$ to maintain consistency of units, since  $I_\textmd{ref}$ is derived from primary voltage and resistance standards using the conventional 1990 values $K_{\text{J-90}}$ and $R_{\text{K-90}}$  for the Josephson and von Klitzing constants respectively \cite{zimmerman1998primer}. The normalised difference between $e_{\text{90}}$ and the latest SI (CODATA 2014) value of $e$ is $(e_{\text{90}}-e)/e_{\text{90}} = -8.06 \times 10^{-8}$ \cite{mohr2016codata}, so consistency of unit systems is an important consideration as the total measurement uncertainty approaches the $0.1$~ppm level. The detailed breakdown of the uncertainty budget for the measurement of $I_\textmd{P}$ is shown in Table~\ref{tab:1}. More information about the precision measurement technique, including a detailed description of each term in Table~\ref{tab:1}, is given in the supplementary information \cite{suppinfo}.

We define the plateau as the region where the fit to the thermal model deviates from the true $ef$ value by less than $0.03$~ppm. We show these $8$ data points on the plateau in the inset of Fig.~\ref{fig:HighResMeas}(a). We performed an additional statistical test, detailed in the supplementary information \cite{suppinfo}, to verify that the scatter of the selected data points is consistent with the data being drawn from the same distribution - in other words, that there is no structure on the plateau within our experimental resolution \cite{giblin2017robust}. Averaging these $N_{\text{P}}=8$ points, we obtain $\Delta I_{\text{P}}=-0.013$~ppm, with standard deviation $\sigma(\Delta I_{\text{P}}) = 0.672$~ppm. We take the error of the mean over the data points as the relative statistical (type A) uncertainty for the measurement of the pumped current, $U_\textmd{A}=\sigma(\Delta I_{\text{P}}) / \sqrt{N_{\text{P}}}$ = 0.237 ppm. The total relative measurement uncertainty of the pumped current, $U_\textmd{total}=0.31$ ppm, is given by the root-sum-square of the eight terms listed in Table~\ref{tab:1}, of which $U_\textmd{A}$ is the largest. Thus the pumped current averaged over the plateau can be expressed as $\Delta I_{\textmd{P}} = (-0.013 \pm 0.31)$ ppm. To verify the robustness of our device, we also carried out another high-accuracy scan by stepping $V_{\textmd{B1}}$. In this scan, we find the pumped current averaged over the plateau, shown in the inset of Fig.~\ref{fig:HighResMeas}(b), to be $\Delta I_{\textmd{P}} = (-0.257 \pm 0.27)$ ppm. The deviation of the pumped current from $ef$ is within the measurement uncertainty, and represents the most accurate measurement to date on a silicon SE pump. This work, along with the previous high-accuracy study of silicon devices in the decay-cascade regime \cite{NTT2016}, indicates that the silicon-based single-electron pump can lead to a more practical and transferable quantum standard of electrical current.
\begin{table}[]
	\centering
	\caption{\label{tab:1} Breakdown of the uncertainty budget for the measurements of the pumped current $I_{\textmd{P}}$ in Fig.~\ref{fig:HighResMeas}. All reported uncertainties are dimensionless $1\sigma$ relative uncertainties.}
	\begin{tabular}{|l|l|l|ll}
		\cline{1-3}
		& PL plateau & B1 plateau &  &  \\ \cline{1-3}
		1. Voltmeter calibration (type A)	& 0.01 ppm      & 0.01 ppm      &  &  \\ \cline{1-3}
		2. Voltmeter linearity	(type B)& 0.03 ppm      & 0.03 ppm      &  &  \\ \cline{1-3}
		3. Voltmeter drift (type B) & 0.068 ppm& 0.068 ppm&  &  \\ \cline{1-3}
		& & & & \\ \cline{1-3}
		4. 1 G$\Omega$ calibration (type B)& 0.1 ppm     & 0.1 ppm&  &  \\ \cline{1-3}
		5. 1 G$\Omega$ drift (type B)& 0.01 ppm& 0.01 ppm&  &  \\ \cline{1-3}
		6. 1 G$\Omega$ calibration (Type A)& 0.15 ppm& 0.07 ppm&  &  \\ \cline{1-3}
		& & & & \\ \cline{1-3}
		7. $I_{\textmd{null}}$	(type B)& 0.01 ppm& 0.01 ppm&  &  \\ \cline{1-3}
		8. $I_{\textmd{P}}$ (type A)& 0.237 ppm& 0.229 ppm&  &  \\ \cline{1-3}
		& & & & \\ \cline{1-3}
		Total	& 0.31 ppm& 0.27 ppm&  &  \\ \cline{1-3}
	\end{tabular}
\end{table}

\section{Summary and Outlook}

Despite severe heating in the electron reservoir, the electron pump presented in this work generated a pumped current equal to $ef$ within the $\sim 0.3$~ppm measurement uncertainty at $f=1$~GHz. Furthermore, fitting the data to a thermal-capture model indicates a theoretical lower bound for the pumping error of 4 ppb at the center of the first current plateau. This suggests that our pump may satisfy the stringent accuracy requirements for a metrological current source \cite{Metrologia2000,keller1996accuracy,PekolaRevModPhys}. In addition, the fact that strong magnetic fields or tailored waveform drives are not required for the accurate operation of our pump, could greatly simplify the experimental implementation of the new standard of electrical current.  
 
 Note that, by adopting a three-waveform pumping scheme \cite{tanttu2016three}, one can potentially reduce the reservoir electron temperature and hence significantly improve the accuracy of our pump in the thermal regime. The three waveform scheme may reduce the rf amplitude required to pump the electrons and mitigate the drive-induced heating in the source reservoir.

\begin{acknowledgments}
We thank T. Tanttu and K. W. Chan for enlightening discussions and valuable feedback on the manuscript. This work is supported by Australian Research Council (Grant No. DP120104710 and DP160104923), the U.S. Army Research Office (W911NF-13-1-0024), 
the Academy of Finland through its Centre of Excellence Programme (Grant No. 284621), the project EMPIR 15SIB08 e-SI-Amp, and the UK department for Business, Innovation, and Skills. This project has received funding from the EMPIR programme co-financed by the Participating States and from the European Union’s Horizon 2020 research and innovation programme. We acknowledge the Australian National Fabrication Facility for the support in device manufacturing. A.R. acknowledges support from the European Union's Horizon 2020 research and innovation programme under the Marie Sklodowska-Curie grant agreement No 654712 (SINHOPSI).
\end{acknowledgments}

\bibliography{ChargePumpBibliography}

\begin{thebibliography}{50}%
\makeatletter
\providecommand \@ifxundefined [1]{%
 \@ifx{#1\undefined}
}%
\providecommand \@ifnum [1]{%
 \ifnum #1\expandafter \@firstoftwo
 \else \expandafter \@secondoftwo
 \fi
}%
\providecommand \@ifx [1]{%
 \ifx #1\expandafter \@firstoftwo
 \else \expandafter \@secondoftwo
 \fi
}%
\providecommand \natexlab [1]{#1}%
\providecommand \enquote  [1]{``#1''}%
\providecommand \bibnamefont  [1]{#1}%
\providecommand \bibfnamefont [1]{#1}%
\providecommand \citenamefont [1]{#1}%
\providecommand \href@noop [0]{\@secondoftwo}%
\providecommand \href [0]{\begingroup \@sanitize@url \@href}%
\providecommand \@href[1]{\@@startlink{#1}\@@href}%
\providecommand \@@href[1]{\endgroup#1\@@endlink}%
\providecommand \@sanitize@url [0]{\catcode `\\12\catcode `\$12\catcode
  `\&12\catcode `\#12\catcode `\^12\catcode `\_12\catcode `\%12\relax}%
\providecommand \@@startlink[1]{}%
\providecommand \@@endlink[0]{}%
\providecommand \url  [0]{\begingroup\@sanitize@url \@url }%
\providecommand \@url [1]{\endgroup\@href {#1}{\urlprefix }}%
\providecommand \urlprefix  [0]{URL }%
\providecommand \Eprint [0]{\href }%
\providecommand \doibase [0]{http://dx.doi.org/}%
\providecommand \selectlanguage [0]{\@gobble}%
\providecommand \bibinfo  [0]{\@secondoftwo}%
\providecommand \bibfield  [0]{\@secondoftwo}%
\providecommand \translation [1]{[#1]}%
\providecommand \BibitemOpen [0]{}%
\providecommand \bibitemStop [0]{}%
\providecommand \bibitemNoStop [0]{.\EOS\space}%
\providecommand \EOS [0]{\spacefactor3000\relax}%
\providecommand \BibitemShut  [1]{\csname bibitem#1\endcsname}%
\let\auto@bib@innerbib\@empty
\bibitem [{\citenamefont {Pekola}\ \emph {et~al.}(2013)\citenamefont {Pekola},
  \citenamefont {Saira}, \citenamefont {Maisi}, \citenamefont {Kemppinen},
  \citenamefont {M\"ott\"onen}, \citenamefont {Pashkin},\ and\ \citenamefont
  {Averin}}]{PekolaRevModPhys}%
  \BibitemOpen
  \bibfield  {author} {\bibinfo {author} {\bibfnamefont {Jukka~P.}\
  \bibnamefont {Pekola}}, \bibinfo {author} {\bibfnamefont {Olli-Pentti}\
  \bibnamefont {Saira}}, \bibinfo {author} {\bibfnamefont {Ville~F.}\
  \bibnamefont {Maisi}}, \bibinfo {author} {\bibfnamefont {Antti}\ \bibnamefont
  {Kemppinen}}, \bibinfo {author} {\bibfnamefont {Mikko}\ \bibnamefont
  {M\"ott\"onen}}, \bibinfo {author} {\bibfnamefont {Yuri~A.}\ \bibnamefont
  {Pashkin}}, \ and\ \bibinfo {author} {\bibfnamefont {Dmitri~V.}\ \bibnamefont
  {Averin}},\ }\bibfield  {title} {\enquote {\bibinfo {title} {Single-electron
  current sources: Toward a refined definition of the ampere},}\ }\href
  {\doibase 10.1103/RevModPhys.85.1421} {\bibfield  {journal} {\bibinfo
  {journal} {Rev. Mod. Phys.}\ }\textbf {\bibinfo {volume} {85}},\ \bibinfo
  {pages} {1421--1472} (\bibinfo {year} {2013})}\BibitemShut {NoStop}%
\bibitem [{\citenamefont {Kaestner}\ and\ \citenamefont
  {Kashcheyevs}(2015)}]{SlavaProgPhys}%
  \BibitemOpen
  \bibfield  {author} {\bibinfo {author} {\bibfnamefont {Bernd}\ \bibnamefont
  {Kaestner}}\ and\ \bibinfo {author} {\bibfnamefont {Vyacheslavs}\
  \bibnamefont {Kashcheyevs}},\ }\bibfield  {title} {\enquote {\bibinfo {title}
  {Non-adiabatic quantized charge pumping with tunable-barrier quantum dots: a
  review of current progress},}\ }\href
  {http://stacks.iop.org/0034-4885/78/i=10/a=103901} {\bibfield  {journal}
  {\bibinfo  {journal} {Rep. Prog. Phys.}\ }\textbf {\bibinfo {volume} {78}},\
  \bibinfo {pages} {103901} (\bibinfo {year} {2015})}\BibitemShut {NoStop}%
\bibitem [{\citenamefont {Kataoka}\ \emph {et~al.}(2016)\citenamefont
  {Kataoka}, \citenamefont {Johnson}, \citenamefont {Emary}, \citenamefont
  {See}, \citenamefont {Griffiths}, \citenamefont {Jones}, \citenamefont
  {Farrer}, \citenamefont {Ritchie}, \citenamefont {Pepper},\ and\
  \citenamefont {Janssen}}]{TimeofFlight}%
  \BibitemOpen
  \bibfield  {author} {\bibinfo {author} {\bibfnamefont {M.}~\bibnamefont
  {Kataoka}}, \bibinfo {author} {\bibfnamefont {N.}~\bibnamefont {Johnson}},
  \bibinfo {author} {\bibfnamefont {C.}~\bibnamefont {Emary}}, \bibinfo
  {author} {\bibfnamefont {P.}~\bibnamefont {See}}, \bibinfo {author}
  {\bibfnamefont {J.~P.}\ \bibnamefont {Griffiths}}, \bibinfo {author}
  {\bibfnamefont {G.~A.~C.}\ \bibnamefont {Jones}}, \bibinfo {author}
  {\bibfnamefont {I.}~\bibnamefont {Farrer}}, \bibinfo {author} {\bibfnamefont
  {D.~A.}\ \bibnamefont {Ritchie}}, \bibinfo {author} {\bibfnamefont
  {M.}~\bibnamefont {Pepper}}, \ and\ \bibinfo {author} {\bibfnamefont {T.~J.
  B.~M.}\ \bibnamefont {Janssen}},\ }\bibfield  {title} {\enquote {\bibinfo
  {title} {Time-of-flight measurements of single-electron wave packets in
  quantum hall edge states},}\ }\href {\doibase 10.1103/PhysRevLett.116.126803}
  {\bibfield  {journal} {\bibinfo  {journal} {Phys. Rev. Lett.}\ }\textbf
  {\bibinfo {volume} {116}},\ \bibinfo {pages} {126803} (\bibinfo {year}
  {2016})}\BibitemShut {NoStop}%
\bibitem [{\citenamefont {Fletcher}\ \emph {et~al.}(2013)\citenamefont
  {Fletcher}, \citenamefont {See}, \citenamefont {Howe}, \citenamefont
  {Pepper}, \citenamefont {Giblin}, \citenamefont {Griffiths}, \citenamefont
  {Jones}, \citenamefont {Farrer}, \citenamefont {Ritchie}, \citenamefont
  {Janssen},\ and\ \citenamefont {Kataoka}}]{ClockedEmission}%
  \BibitemOpen
  \bibfield  {author} {\bibinfo {author} {\bibfnamefont {J.~D.}\ \bibnamefont
  {Fletcher}}, \bibinfo {author} {\bibfnamefont {P.}~\bibnamefont {See}},
  \bibinfo {author} {\bibfnamefont {H.}~\bibnamefont {Howe}}, \bibinfo {author}
  {\bibfnamefont {M.}~\bibnamefont {Pepper}}, \bibinfo {author} {\bibfnamefont
  {S.~P.}\ \bibnamefont {Giblin}}, \bibinfo {author} {\bibfnamefont {J.~P.}\
  \bibnamefont {Griffiths}}, \bibinfo {author} {\bibfnamefont {G.~A.~C.}\
  \bibnamefont {Jones}}, \bibinfo {author} {\bibfnamefont {I.}~\bibnamefont
  {Farrer}}, \bibinfo {author} {\bibfnamefont {D.~A.}\ \bibnamefont {Ritchie}},
  \bibinfo {author} {\bibfnamefont {T.~J. B.~M.}\ \bibnamefont {Janssen}}, \
  and\ \bibinfo {author} {\bibfnamefont {M.}~\bibnamefont {Kataoka}},\
  }\bibfield  {title} {\enquote {\bibinfo {title} {Clock-controlled emission of
  single-electron wave packets in a solid-state circuit},}\ }\href {\doibase
  10.1103/PhysRevLett.111.216807} {\bibfield  {journal} {\bibinfo  {journal}
  {Phys. Rev. Lett.}\ }\textbf {\bibinfo {volume} {111}},\ \bibinfo {pages}
  {216807} (\bibinfo {year} {2013})}\BibitemShut {NoStop}%
\bibitem [{\citenamefont {Ubbelohde}\ \emph {et~al.}(2015)\citenamefont
  {Ubbelohde}, \citenamefont {Hohls}, \citenamefont {Kashcheyevs},
  \citenamefont {Wagner}, \citenamefont {Fricke}, \citenamefont {K{\"a}stner},
  \citenamefont {Pierz}, \citenamefont {Schumacher},\ and\ \citenamefont
  {Haug}}]{ubbelohde2015partitioning}%
  \BibitemOpen
  \bibfield  {author} {\bibinfo {author} {\bibfnamefont {Niels}\ \bibnamefont
  {Ubbelohde}}, \bibinfo {author} {\bibfnamefont {Frank}\ \bibnamefont
  {Hohls}}, \bibinfo {author} {\bibfnamefont {Vyacheslavs}\ \bibnamefont
  {Kashcheyevs}}, \bibinfo {author} {\bibfnamefont {Timo}\ \bibnamefont
  {Wagner}}, \bibinfo {author} {\bibfnamefont {Lukas}\ \bibnamefont {Fricke}},
  \bibinfo {author} {\bibfnamefont {Bernd}\ \bibnamefont {K{\"a}stner}},
  \bibinfo {author} {\bibfnamefont {Klaus}\ \bibnamefont {Pierz}}, \bibinfo
  {author} {\bibfnamefont {Hans~W}\ \bibnamefont {Schumacher}}, \ and\ \bibinfo
  {author} {\bibfnamefont {Rolf~J}\ \bibnamefont {Haug}},\ }\bibfield  {title}
  {\enquote {\bibinfo {title} {Partitioning of on-demand electron pairs},}\
  }\href@noop {} {\bibfield  {journal} {\bibinfo  {journal} {Nature
  nanotechnology}\ }\textbf {\bibinfo {volume} {10}},\ \bibinfo {pages}
  {46--49} (\bibinfo {year} {2015})}\BibitemShut {NoStop}%
\bibitem [{\citenamefont {Piquemal}\ and\ \citenamefont
  {Genevès}(2000)}]{Metrologia2000}%
  \BibitemOpen
  \bibfield  {author} {\bibinfo {author} {\bibfnamefont {F.}~\bibnamefont
  {Piquemal}}\ and\ \bibinfo {author} {\bibfnamefont {G.}~\bibnamefont
  {Genevès}},\ }\bibfield  {title} {\enquote {\bibinfo {title} {Argument for a
  direct realization of the quantum metrological triangle},}\ }\href
  {http://stacks.iop.org/0026-1394/37/i=3/a=4} {\bibfield  {journal} {\bibinfo
  {journal} {Metrologia}\ }\textbf {\bibinfo {volume} {37}},\ \bibinfo {pages}
  {207} (\bibinfo {year} {2000})}\BibitemShut {NoStop}%
\bibitem [{\citenamefont {Keller}\ \emph {et~al.}(1996)\citenamefont {Keller},
  \citenamefont {Martinis}, \citenamefont {Zimmerman},\ and\ \citenamefont
  {Steinbach}}]{keller1996accuracy}%
  \BibitemOpen
  \bibfield  {author} {\bibinfo {author} {\bibfnamefont {Mark~W.}\ \bibnamefont
  {Keller}}, \bibinfo {author} {\bibfnamefont {John~M.}\ \bibnamefont
  {Martinis}}, \bibinfo {author} {\bibfnamefont {Neil~M.}\ \bibnamefont
  {Zimmerman}}, \ and\ \bibinfo {author} {\bibfnamefont {Andrew~H.}\
  \bibnamefont {Steinbach}},\ }\bibfield  {title} {\enquote {\bibinfo {title}
  {Accuracy of electron counting using a 7-junction electron pump},}\
  }\href@noop {} {\bibfield  {journal} {\bibinfo  {journal} {Appl. Phys.
  Lett.}\ }\textbf {\bibinfo {volume} {69}},\ \bibinfo {pages} {1804--1806}
  (\bibinfo {year} {1996})}\BibitemShut {NoStop}%
\bibitem [{\citenamefont {Keller}\ \emph {et~al.}(1999)\citenamefont {Keller},
  \citenamefont {Eichenberger}, \citenamefont {Martinis},\ and\ \citenamefont
  {Zimmerman}}]{keller1999capacitance}%
  \BibitemOpen
  \bibfield  {author} {\bibinfo {author} {\bibfnamefont {Mark~W.}\ \bibnamefont
  {Keller}}, \bibinfo {author} {\bibfnamefont {Ali~L.}\ \bibnamefont
  {Eichenberger}}, \bibinfo {author} {\bibfnamefont {John~M.}\ \bibnamefont
  {Martinis}}, \ and\ \bibinfo {author} {\bibfnamefont {Neil~M.}\ \bibnamefont
  {Zimmerman}},\ }\bibfield  {title} {\enquote {\bibinfo {title} {A capacitance
  standard based on counting electrons},}\ }\href@noop {} {\bibfield  {journal}
  {\bibinfo  {journal} {Science}\ }\textbf {\bibinfo {volume} {285}},\ \bibinfo
  {pages} {1706--1709} (\bibinfo {year} {1999})}\BibitemShut {NoStop}%
\bibitem [{\citenamefont {Shilton}\ \emph {et~al.}(1996)\citenamefont
  {Shilton}, \citenamefont {Talyanskii}, \citenamefont {Pepper}, \citenamefont
  {Ritchie}, \citenamefont {Frost}, \citenamefont {Ford}, \citenamefont
  {Smith},\ and\ \citenamefont {Jones}}]{shilton1996high}%
  \BibitemOpen
  \bibfield  {author} {\bibinfo {author} {\bibfnamefont {J.M.}\ \bibnamefont
  {Shilton}}, \bibinfo {author} {\bibfnamefont {V.I.}\ \bibnamefont
  {Talyanskii}}, \bibinfo {author} {\bibfnamefont {M.}~\bibnamefont {Pepper}},
  \bibinfo {author} {\bibfnamefont {D.A.}\ \bibnamefont {Ritchie}}, \bibinfo
  {author} {\bibfnamefont {J.E.F.}\ \bibnamefont {Frost}}, \bibinfo {author}
  {\bibfnamefont {C.J.B.}\ \bibnamefont {Ford}}, \bibinfo {author}
  {\bibfnamefont {C.G.}\ \bibnamefont {Smith}}, \ and\ \bibinfo {author}
  {\bibfnamefont {G.A.C.}\ \bibnamefont {Jones}},\ }\bibfield  {title}
  {\enquote {\bibinfo {title} {High-frequency single-electron transport in a
  quasi-one-dimensional gaas channel induced by surface acoustic waves},}\
  }\href@noop {} {\bibfield  {journal} {\bibinfo  {journal} {J. Phys. Condens.
  Matter}\ }\textbf {\bibinfo {volume} {8}},\ \bibinfo {pages} {L531} (\bibinfo
  {year} {1996})}\BibitemShut {NoStop}%
\bibitem [{\citenamefont {Talyanskii}\ \emph {et~al.}(1997)\citenamefont
  {Talyanskii}, \citenamefont {Shilton}, \citenamefont {Pepper}, \citenamefont
  {Smith}, \citenamefont {Ford}, \citenamefont {Linfield}, \citenamefont
  {Ritchie},\ and\ \citenamefont {Jones}}]{talyanskii1997single}%
  \BibitemOpen
  \bibfield  {author} {\bibinfo {author} {\bibfnamefont {V.I.}\ \bibnamefont
  {Talyanskii}}, \bibinfo {author} {\bibfnamefont {J.M.}\ \bibnamefont
  {Shilton}}, \bibinfo {author} {\bibfnamefont {M.}~\bibnamefont {Pepper}},
  \bibinfo {author} {\bibfnamefont {C.G.}\ \bibnamefont {Smith}}, \bibinfo
  {author} {\bibfnamefont {C.J.B.}\ \bibnamefont {Ford}}, \bibinfo {author}
  {\bibfnamefont {E.H.}\ \bibnamefont {Linfield}}, \bibinfo {author}
  {\bibfnamefont {D.A.}\ \bibnamefont {Ritchie}}, \ and\ \bibinfo {author}
  {\bibfnamefont {G.A.C.}\ \bibnamefont {Jones}},\ }\bibfield  {title}
  {\enquote {\bibinfo {title} {Single-electron transport in a one-dimensional
  channel by high-frequency surface acoustic waves},}\ }\href@noop {}
  {\bibfield  {journal} {\bibinfo  {journal} {Phys. Rev. B}\ }\textbf {\bibinfo
  {volume} {56}},\ \bibinfo {pages} {15180} (\bibinfo {year}
  {1997})}\BibitemShut {NoStop}%
\bibitem [{\citenamefont {Niskanen}\ \emph {et~al.}(2005)\citenamefont
  {Niskanen}, \citenamefont {Kivioja}, \citenamefont {Sepp{\"a}},\ and\
  \citenamefont {Pekola}}]{niskanen2005evidence}%
  \BibitemOpen
  \bibfield  {author} {\bibinfo {author} {\bibfnamefont {Antti~O.}\
  \bibnamefont {Niskanen}}, \bibinfo {author} {\bibfnamefont {Jani~M.}\
  \bibnamefont {Kivioja}}, \bibinfo {author} {\bibfnamefont {Heikki}\
  \bibnamefont {Sepp{\"a}}}, \ and\ \bibinfo {author} {\bibfnamefont
  {Jukka~P.}\ \bibnamefont {Pekola}},\ }\bibfield  {title} {\enquote {\bibinfo
  {title} {Evidence of cooper-pair pumping with combined flux and voltage
  control},}\ }\href@noop {} {\bibfield  {journal} {\bibinfo  {journal} {Phys.
  Rev. B}\ }\textbf {\bibinfo {volume} {71}},\ \bibinfo {pages} {012513}
  (\bibinfo {year} {2005})}\BibitemShut {NoStop}%
\bibitem [{\citenamefont {Vartiainen}\ \emph {et~al.}(2007)\citenamefont
  {Vartiainen}, \citenamefont {M{\"o}tt{\"o}nen}, \citenamefont {Pekola},\ and\
  \citenamefont {Kemppinen}}]{vartiainen2007nanoampere}%
  \BibitemOpen
  \bibfield  {author} {\bibinfo {author} {\bibfnamefont {Juha~J.}\ \bibnamefont
  {Vartiainen}}, \bibinfo {author} {\bibfnamefont {Mikko}\ \bibnamefont
  {M{\"o}tt{\"o}nen}}, \bibinfo {author} {\bibfnamefont {Jukka~P.}\
  \bibnamefont {Pekola}}, \ and\ \bibinfo {author} {\bibfnamefont {Antti}\
  \bibnamefont {Kemppinen}},\ }\bibfield  {title} {\enquote {\bibinfo {title}
  {Nanoampere pumping of cooper pairs},}\ }\href@noop {} {\bibfield  {journal}
  {\bibinfo  {journal} {Appl. Phys. Lett.}\ }\textbf {\bibinfo {volume} {90}},\
  \bibinfo {pages} {082102} (\bibinfo {year} {2007})}\BibitemShut {NoStop}%
\bibitem [{\citenamefont {M{\"o}tt{\"o}nen}\ \emph {et~al.}(2008)\citenamefont
  {M{\"o}tt{\"o}nen}, \citenamefont {Vartiainen},\ and\ \citenamefont
  {Pekola}}]{mottonen2008experimental}%
  \BibitemOpen
  \bibfield  {author} {\bibinfo {author} {\bibfnamefont {Mikko}\ \bibnamefont
  {M{\"o}tt{\"o}nen}}, \bibinfo {author} {\bibfnamefont {Juha~J.}\ \bibnamefont
  {Vartiainen}}, \ and\ \bibinfo {author} {\bibfnamefont {Jukka~P.}\
  \bibnamefont {Pekola}},\ }\bibfield  {title} {\enquote {\bibinfo {title}
  {Experimental determination of the berry phase in a superconducting charge
  pump},}\ }\href@noop {} {\bibfield  {journal} {\bibinfo  {journal} {Physical
  review letters}\ }\textbf {\bibinfo {volume} {100}},\ \bibinfo {pages}
  {177201} (\bibinfo {year} {2008})}\BibitemShut {NoStop}%
\bibitem [{\citenamefont {Pekola}\ \emph {et~al.}(2008)\citenamefont {Pekola},
  \citenamefont {Vartiainen}, \citenamefont {M{\"o}tt{\"o}nen}, \citenamefont
  {Saira}, \citenamefont {Meschke},\ and\ \citenamefont
  {Averin}}]{pekola2008hybrid}%
  \BibitemOpen
  \bibfield  {author} {\bibinfo {author} {\bibfnamefont {Jukka~P.}\
  \bibnamefont {Pekola}}, \bibinfo {author} {\bibfnamefont {Juha~J.}\
  \bibnamefont {Vartiainen}}, \bibinfo {author} {\bibfnamefont {Mikko}\
  \bibnamefont {M{\"o}tt{\"o}nen}}, \bibinfo {author} {\bibfnamefont
  {Olli-Pentti}\ \bibnamefont {Saira}}, \bibinfo {author} {\bibfnamefont
  {Matthias}\ \bibnamefont {Meschke}}, \ and\ \bibinfo {author} {\bibfnamefont
  {Dmitri~V.}\ \bibnamefont {Averin}},\ }\bibfield  {title} {\enquote {\bibinfo
  {title} {Hybrid single-electron transistor as a source of quantized electric
  current},}\ }\href@noop {} {\bibfield  {journal} {\bibinfo  {journal} {Nat.
  Phys.}\ }\textbf {\bibinfo {volume} {4}},\ \bibinfo {pages} {120--124}
  (\bibinfo {year} {2008})}\BibitemShut {NoStop}%
\bibitem [{\citenamefont {Blumenthal}\ \emph {et~al.}(2007)\citenamefont
  {Blumenthal}, \citenamefont {Kaestner}, \citenamefont {Li}, \citenamefont
  {Giblin}, \citenamefont {Janssen}, \citenamefont {Pepper}, \citenamefont
  {Anderson}, \citenamefont {Jones},\ and\ \citenamefont
  {Ritchie}}]{blumenthal2007gigahertz}%
  \BibitemOpen
  \bibfield  {author} {\bibinfo {author} {\bibfnamefont {M.D.}\ \bibnamefont
  {Blumenthal}}, \bibinfo {author} {\bibfnamefont {B.}~\bibnamefont
  {Kaestner}}, \bibinfo {author} {\bibfnamefont {L.}~\bibnamefont {Li}},
  \bibinfo {author} {\bibfnamefont {S.}~\bibnamefont {Giblin}}, \bibinfo
  {author} {\bibfnamefont {T.J.B.M.}\ \bibnamefont {Janssen}}, \bibinfo
  {author} {\bibfnamefont {M.}~\bibnamefont {Pepper}}, \bibinfo {author}
  {\bibfnamefont {D.}~\bibnamefont {Anderson}}, \bibinfo {author}
  {\bibfnamefont {G.}~\bibnamefont {Jones}}, \ and\ \bibinfo {author}
  {\bibfnamefont {D.A.}\ \bibnamefont {Ritchie}},\ }\bibfield  {title}
  {\enquote {\bibinfo {title} {Gigahertz quantized charge pumping},}\
  }\href@noop {} {\bibfield  {journal} {\bibinfo  {journal} {Nat. Phys.}\
  }\textbf {\bibinfo {volume} {3}},\ \bibinfo {pages} {343--347} (\bibinfo
  {year} {2007})}\BibitemShut {NoStop}%
\bibitem [{\citenamefont {Jehl}\ \emph {et~al.}(2013)\citenamefont {Jehl} \emph
  {et~al.}}]{jehl2013hybrid}%
  \BibitemOpen
  \bibfield  {author} {\bibinfo {author} {\bibfnamefont {X.}~\bibnamefont
  {Jehl}} \emph {et~al.},\ }\bibfield  {title} {\enquote {\bibinfo {title}
  {Hybrid metal-semiconductor electron pump for quantum metrology},}\
  }\href@noop {} {\bibfield  {journal} {\bibinfo  {journal} {Phys. Rev. X}\
  }\textbf {\bibinfo {volume} {3}},\ \bibinfo {pages} {021012} (\bibinfo {year}
  {2013})}\BibitemShut {NoStop}%
\bibitem [{\citenamefont {Giblin}\ \emph {et~al.}(2012)\citenamefont {Giblin},
  \citenamefont {Kataoka}, \citenamefont {Fletcher}, \citenamefont {See},
  \citenamefont {Janssen}, \citenamefont {Griffiths}, \citenamefont {Jones},
  \citenamefont {Farrer},\ and\ \citenamefont {Ritchie}}]{NPL2012}%
  \BibitemOpen
  \bibfield  {author} {\bibinfo {author} {\bibfnamefont {S.P.}\ \bibnamefont
  {Giblin}}, \bibinfo {author} {\bibfnamefont {M.}~\bibnamefont {Kataoka}},
  \bibinfo {author} {\bibfnamefont {J.D.}\ \bibnamefont {Fletcher}}, \bibinfo
  {author} {\bibfnamefont {P.}~\bibnamefont {See}}, \bibinfo {author}
  {\bibfnamefont {T.J.B.M.}\ \bibnamefont {Janssen}}, \bibinfo {author}
  {\bibfnamefont {J.P.}\ \bibnamefont {Griffiths}}, \bibinfo {author}
  {\bibfnamefont {G.A.C.}\ \bibnamefont {Jones}}, \bibinfo {author}
  {\bibfnamefont {I.}~\bibnamefont {Farrer}}, \ and\ \bibinfo {author}
  {\bibfnamefont {D.A.}\ \bibnamefont {Ritchie}},\ }\bibfield  {title}
  {\enquote {\bibinfo {title} {Towards a quantum representation of the ampere
  using single electron pumps},}\ }\href@noop {} {\bibfield  {journal}
  {\bibinfo  {journal} {Nat. Commun.}\ }\textbf {\bibinfo {volume} {3}},\
  \bibinfo {pages} {930} (\bibinfo {year} {2012})}\BibitemShut {NoStop}%
\bibitem [{\citenamefont {Stein}\ \emph {et~al.}(2015)\citenamefont {Stein}
  \emph {et~al.}}]{PTB2015}%
  \BibitemOpen
  \bibfield  {author} {\bibinfo {author} {\bibfnamefont {Friederike}\
  \bibnamefont {Stein}} \emph {et~al.},\ }\bibfield  {title} {\enquote
  {\bibinfo {title} {Validation of a quantized-current source with 0.2 ppm
  uncertainty},}\ }\href@noop {} {\bibfield  {journal} {\bibinfo  {journal}
  {Appl. Phys. Lett.}\ }\textbf {\bibinfo {volume} {107}},\ \bibinfo {pages}
  {103501} (\bibinfo {year} {2015})}\BibitemShut {NoStop}%
\bibitem [{\citenamefont {Giblin}\ \emph {et~al.}(2016)\citenamefont {Giblin},
  \citenamefont {See}, \citenamefont {Petrie}, \citenamefont {Janssen},
  \citenamefont {Farrer}, \citenamefont {Griffiths}, \citenamefont {Jones},
  \citenamefont {Ritchie},\ and\ \citenamefont {Kataoka}}]{giblin2016high}%
  \BibitemOpen
  \bibfield  {author} {\bibinfo {author} {\bibfnamefont {S.P.}\ \bibnamefont
  {Giblin}}, \bibinfo {author} {\bibfnamefont {P.}~\bibnamefont {See}},
  \bibinfo {author} {\bibfnamefont {A.}~\bibnamefont {Petrie}}, \bibinfo
  {author} {\bibfnamefont {T.J.B.M.}\ \bibnamefont {Janssen}}, \bibinfo
  {author} {\bibfnamefont {I.}~\bibnamefont {Farrer}}, \bibinfo {author}
  {\bibfnamefont {J.P.}\ \bibnamefont {Griffiths}}, \bibinfo {author}
  {\bibfnamefont {G.A.C.}\ \bibnamefont {Jones}}, \bibinfo {author}
  {\bibfnamefont {D.A.}\ \bibnamefont {Ritchie}}, \ and\ \bibinfo {author}
  {\bibfnamefont {M.}~\bibnamefont {Kataoka}},\ }\bibfield  {title} {\enquote
  {\bibinfo {title} {High-resolution error detection in the capture process of
  a single-electron pump},}\ }\href@noop {} {\bibfield  {journal} {\bibinfo
  {journal} {Appl. Phys. Lett.}\ }\textbf {\bibinfo {volume} {108}},\ \bibinfo
  {pages} {023502} (\bibinfo {year} {2016})}\BibitemShut {NoStop}%
\bibitem [{\citenamefont {Connolly}\ \emph {et~al.}(2013)\citenamefont
  {Connolly}, \citenamefont {Chiu}, \citenamefont {Giblin}, \citenamefont
  {Kataoka}, \citenamefont {Fletcher}, \citenamefont {Chua}, \citenamefont
  {Griffiths}, \citenamefont {Jones}, \citenamefont {Fal'Ko}, \citenamefont
  {Smith},\ and\ \citenamefont {Janssen}}]{connolly2013gigahertz}%
  \BibitemOpen
  \bibfield  {author} {\bibinfo {author} {\bibfnamefont {M.R.}\ \bibnamefont
  {Connolly}}, \bibinfo {author} {\bibfnamefont {K.L.}\ \bibnamefont {Chiu}},
  \bibinfo {author} {\bibfnamefont {S.P.}\ \bibnamefont {Giblin}}, \bibinfo
  {author} {\bibfnamefont {M.}~\bibnamefont {Kataoka}}, \bibinfo {author}
  {\bibfnamefont {J.D.}\ \bibnamefont {Fletcher}}, \bibinfo {author}
  {\bibfnamefont {C.}~\bibnamefont {Chua}}, \bibinfo {author} {\bibfnamefont
  {J.P.}\ \bibnamefont {Griffiths}}, \bibinfo {author} {\bibfnamefont {G.A.C.}\
  \bibnamefont {Jones}}, \bibinfo {author} {\bibfnamefont {V.I.}\ \bibnamefont
  {Fal'Ko}}, \bibinfo {author} {\bibfnamefont {C.G.}\ \bibnamefont {Smith}}, \
  and\ \bibinfo {author} {\bibfnamefont {T.J.B.M.}\ \bibnamefont {Janssen}},\
  }\bibfield  {title} {\enquote {\bibinfo {title} {Gigahertz quantized charge
  pumping in graphene quantum dots},}\ }\href@noop {} {\bibfield  {journal}
  {\bibinfo  {journal} {Nat. Nanotech.}\ }\textbf {\bibinfo {volume} {8}},\
  \bibinfo {pages} {417--420} (\bibinfo {year} {2013})}\BibitemShut {NoStop}%
\bibitem [{\citenamefont {Fujiwara}\ \emph {et~al.}(2004)\citenamefont
  {Fujiwara}, \citenamefont {Zimmerman}, \citenamefont {Ono},\ and\
  \citenamefont {Takahashi}}]{fujiwara2004current}%
  \BibitemOpen
  \bibfield  {author} {\bibinfo {author} {\bibfnamefont {Akira}\ \bibnamefont
  {Fujiwara}}, \bibinfo {author} {\bibfnamefont {Neil~M.}\ \bibnamefont
  {Zimmerman}}, \bibinfo {author} {\bibfnamefont {Yukinori}\ \bibnamefont
  {Ono}}, \ and\ \bibinfo {author} {\bibfnamefont {Yasuo}\ \bibnamefont
  {Takahashi}},\ }\bibfield  {title} {\enquote {\bibinfo {title} {Current
  quantization due to single-electron transfer in si-wire charge-coupled
  devices},}\ }\href@noop {} {\bibfield  {journal} {\bibinfo  {journal} {Appl.
  Phys. Lett.}\ }\textbf {\bibinfo {volume} {84}},\ \bibinfo {pages}
  {1323--1325} (\bibinfo {year} {2004})}\BibitemShut {NoStop}%
\bibitem [{\citenamefont {Fujiwara}\ \emph {et~al.}(2008)\citenamefont
  {Fujiwara}, \citenamefont {Nishiguchi},\ and\ \citenamefont
  {Ono}}]{fujiwara2008nanoampere}%
  \BibitemOpen
  \bibfield  {author} {\bibinfo {author} {\bibfnamefont {Akira}\ \bibnamefont
  {Fujiwara}}, \bibinfo {author} {\bibfnamefont {Katsuhiko}\ \bibnamefont
  {Nishiguchi}}, \ and\ \bibinfo {author} {\bibfnamefont {Yukinori}\
  \bibnamefont {Ono}},\ }\bibfield  {title} {\enquote {\bibinfo {title}
  {Nanoampere charge pump by single-electron ratchet using silicon nanowire
  metal-oxide-semiconductor field-effect transistor},}\ }\href@noop {}
  {\bibfield  {journal} {\bibinfo  {journal} {Appl. Phys. Lett.}\ }\textbf
  {\bibinfo {volume} {92}},\ \bibinfo {pages} {2102} (\bibinfo {year}
  {2008})}\BibitemShut {NoStop}%
\bibitem [{\citenamefont {Ono}\ and\ \citenamefont
  {Takahashi}(2003)}]{ono2003electron}%
  \BibitemOpen
  \bibfield  {author} {\bibinfo {author} {\bibfnamefont {Yukinori}\
  \bibnamefont {Ono}}\ and\ \bibinfo {author} {\bibfnamefont {Yasuo}\
  \bibnamefont {Takahashi}},\ }\bibfield  {title} {\enquote {\bibinfo {title}
  {Electron pump by a combined single-electron/field-effect-transistor
  structure},}\ }\href@noop {} {\bibfield  {journal} {\bibinfo  {journal}
  {Appl. Phys. Lett.}\ }\textbf {\bibinfo {volume} {82}},\ \bibinfo {pages}
  {1221--1223} (\bibinfo {year} {2003})}\BibitemShut {NoStop}%
\bibitem [{\citenamefont {Chan}\ \emph {et~al.}(2011)\citenamefont {Chan},
  \citenamefont {M{\"o}tt{\"o}nen}, \citenamefont {Kemppinen}, \citenamefont
  {Lai}, \citenamefont {Tan}, \citenamefont {Lim},\ and\ \citenamefont
  {Dzurak}}]{chan2011single}%
  \BibitemOpen
  \bibfield  {author} {\bibinfo {author} {\bibfnamefont {K.W.}\ \bibnamefont
  {Chan}}, \bibinfo {author} {\bibfnamefont {M.}~\bibnamefont
  {M{\"o}tt{\"o}nen}}, \bibinfo {author} {\bibfnamefont {A.}~\bibnamefont
  {Kemppinen}}, \bibinfo {author} {\bibfnamefont {N.S.}\ \bibnamefont {Lai}},
  \bibinfo {author} {\bibfnamefont {K.Y.}\ \bibnamefont {Tan}}, \bibinfo
  {author} {\bibfnamefont {W.H.}\ \bibnamefont {Lim}}, \ and\ \bibinfo {author}
  {\bibfnamefont {A.S.}\ \bibnamefont {Dzurak}},\ }\bibfield  {title} {\enquote
  {\bibinfo {title} {Single-electron shuttle based on a silicon quantum dot},}\
  }\href@noop {} {\bibfield  {journal} {\bibinfo  {journal} {Appl. Phys.
  Lett.}\ }\textbf {\bibinfo {volume} {98}},\ \bibinfo {pages} {212103}
  (\bibinfo {year} {2011})}\BibitemShut {NoStop}%
\bibitem [{\citenamefont {Rossi}\ \emph {et~al.}(2014)\citenamefont {Rossi},
  \citenamefont {Tanttu}, \citenamefont {Tan}, \citenamefont {Iisakka},
  \citenamefont {Zhao}, \citenamefont {Chan}, \citenamefont {Tettamanzi},
  \citenamefont {Rogge}, \citenamefont {Dzurak},\ and\ \citenamefont
  {M\"ott\"onen}}]{Ale2014}%
  \BibitemOpen
  \bibfield  {author} {\bibinfo {author} {\bibfnamefont {Alessandro}\
  \bibnamefont {Rossi}}, \bibinfo {author} {\bibfnamefont {Tuomo}\ \bibnamefont
  {Tanttu}}, \bibinfo {author} {\bibfnamefont {Kuan~Yen}\ \bibnamefont {Tan}},
  \bibinfo {author} {\bibfnamefont {Ilkka}\ \bibnamefont {Iisakka}}, \bibinfo
  {author} {\bibfnamefont {Ruichen}\ \bibnamefont {Zhao}}, \bibinfo {author}
  {\bibfnamefont {Kok~Wai}\ \bibnamefont {Chan}}, \bibinfo {author}
  {\bibfnamefont {Giuseppe~C}\ \bibnamefont {Tettamanzi}}, \bibinfo {author}
  {\bibfnamefont {Sven}\ \bibnamefont {Rogge}}, \bibinfo {author}
  {\bibfnamefont {Andrew~S.}\ \bibnamefont {Dzurak}}, \ and\ \bibinfo {author}
  {\bibfnamefont {Mikko}\ \bibnamefont {M\"ott\"onen}},\ }\bibfield  {title}
  {\enquote {\bibinfo {title} {An accurate single-electron pump based on a
  highly tunable silicon quantum dot},}\ }\href@noop {} {\bibfield  {journal}
  {\bibinfo  {journal} {Nano Lett.}\ }\textbf {\bibinfo {volume} {14}},\
  \bibinfo {pages} {3405--3411} (\bibinfo {year} {2014})}\BibitemShut {NoStop}%
\bibitem [{\citenamefont {Tanttu}\ \emph {et~al.}(2016)\citenamefont {Tanttu},
  \citenamefont {Rossi}, \citenamefont {Tan}, \citenamefont {M{\"a}kinen},
  \citenamefont {Chan}, \citenamefont {Dzurak},\ and\ \citenamefont
  {M{\"o}tt{\"o}nen}}]{tanttu2016three}%
  \BibitemOpen
  \bibfield  {author} {\bibinfo {author} {\bibfnamefont {Tuomo}\ \bibnamefont
  {Tanttu}}, \bibinfo {author} {\bibfnamefont {Alessandro}\ \bibnamefont
  {Rossi}}, \bibinfo {author} {\bibfnamefont {Kuan~Yen}\ \bibnamefont {Tan}},
  \bibinfo {author} {\bibfnamefont {Akseli}\ \bibnamefont {M{\"a}kinen}},
  \bibinfo {author} {\bibfnamefont {Kok~Wai}\ \bibnamefont {Chan}}, \bibinfo
  {author} {\bibfnamefont {Andrew~S}\ \bibnamefont {Dzurak}}, \ and\ \bibinfo
  {author} {\bibfnamefont {Mikko}\ \bibnamefont {M{\"o}tt{\"o}nen}},\
  }\bibfield  {title} {\enquote {\bibinfo {title} {Three-waveform bidirectional
  pumping of single electrons with a silicon quantum dot},}\ }\href@noop {}
  {\bibfield  {journal} {\bibinfo  {journal} {Scientific Reports}\ }\textbf
  {\bibinfo {volume} {6}},\ \bibinfo {pages} {36381} (\bibinfo {year}
  {2016})}\BibitemShut {NoStop}%
\bibitem [{\citenamefont {Yamahata}\ \emph {et~al.}(2016)\citenamefont
  {Yamahata}, \citenamefont {Giblin}, \citenamefont {Kataoka}, \citenamefont
  {Karasawa},\ and\ \citenamefont {Fujiwara}}]{NTT2016}%
  \BibitemOpen
  \bibfield  {author} {\bibinfo {author} {\bibfnamefont {Gento}\ \bibnamefont
  {Yamahata}}, \bibinfo {author} {\bibfnamefont {Stephen~P.}\ \bibnamefont
  {Giblin}}, \bibinfo {author} {\bibfnamefont {Masaya}\ \bibnamefont
  {Kataoka}}, \bibinfo {author} {\bibfnamefont {Takeshi}\ \bibnamefont
  {Karasawa}}, \ and\ \bibinfo {author} {\bibfnamefont {Akira}\ \bibnamefont
  {Fujiwara}},\ }\bibfield  {title} {\enquote {\bibinfo {title} {Gigahertz
  single-electron pumping in silicon with an accuracy better than 9.2 parts in
  107},}\ }\href@noop {} {\bibfield  {journal} {\bibinfo  {journal} {Appl.
  Phys. Lett.}\ }\textbf {\bibinfo {volume} {109}},\ \bibinfo {eid} {013101}
  (\bibinfo {year} {2016})}\BibitemShut {NoStop}%
\bibitem [{\citenamefont {Yamahata}\ \emph {et~al.}(2017)\citenamefont
  {Yamahata}, \citenamefont {Giblin}, \citenamefont {Kataoka}, \citenamefont
  {Karasawa},\ and\ \citenamefont {Fujiwara}}]{yamahata2017high}%
  \BibitemOpen
  \bibfield  {author} {\bibinfo {author} {\bibfnamefont {Gento}\ \bibnamefont
  {Yamahata}}, \bibinfo {author} {\bibfnamefont {Stephen~P}\ \bibnamefont
  {Giblin}}, \bibinfo {author} {\bibfnamefont {Masaya}\ \bibnamefont
  {Kataoka}}, \bibinfo {author} {\bibfnamefont {Takeshi}\ \bibnamefont
  {Karasawa}}, \ and\ \bibinfo {author} {\bibfnamefont {Akira}\ \bibnamefont
  {Fujiwara}},\ }\bibfield  {title} {\enquote {\bibinfo {title} {High-accuracy
  current generation in the nanoampere regime from a silicon single-trap
  electron pump},}\ }\href@noop {} {\bibfield  {journal} {\bibinfo  {journal}
  {Scientific Reports}\ }\textbf {\bibinfo {volume} {7}} (\bibinfo {year}
  {2017})}\BibitemShut {NoStop}%
\bibitem [{\citenamefont {Yamahata}\ \emph
  {et~al.}(2014{\natexlab{a}})\citenamefont {Yamahata}, \citenamefont
  {Nishiguchi},\ and\ \citenamefont {Fujiwara}}]{yamahata2014gigahertz}%
  \BibitemOpen
  \bibfield  {author} {\bibinfo {author} {\bibfnamefont {Gento}\ \bibnamefont
  {Yamahata}}, \bibinfo {author} {\bibfnamefont {Katsuhiko}\ \bibnamefont
  {Nishiguchi}}, \ and\ \bibinfo {author} {\bibfnamefont {Akira}\ \bibnamefont
  {Fujiwara}},\ }\bibfield  {title} {\enquote {\bibinfo {title} {Gigahertz
  single-trap electron pumps in silicon},}\ }\href@noop {} {\bibfield
  {journal} {\bibinfo  {journal} {Nat. Commun.}\ }\textbf {\bibinfo {volume}
  {5}},\ \bibinfo {pages} {5038} (\bibinfo {year}
  {2014}{\natexlab{a}})}\BibitemShut {NoStop}%
\bibitem [{\citenamefont {Tettamanzi}\ \emph {et~al.}(2014)\citenamefont
  {Tettamanzi}, \citenamefont {Wacquez},\ and\ \citenamefont
  {Rogge}}]{tettamanzi2014charge}%
  \BibitemOpen
  \bibfield  {author} {\bibinfo {author} {\bibfnamefont {G.C.}\ \bibnamefont
  {Tettamanzi}}, \bibinfo {author} {\bibfnamefont {Romain}\ \bibnamefont
  {Wacquez}}, \ and\ \bibinfo {author} {\bibfnamefont {S.}~\bibnamefont
  {Rogge}},\ }\bibfield  {title} {\enquote {\bibinfo {title} {Charge pumping
  through a single donor atom},}\ }\href@noop {} {\bibfield  {journal}
  {\bibinfo  {journal} {New J. Phys.}\ }\textbf {\bibinfo {volume} {16}},\
  \bibinfo {pages} {063036} (\bibinfo {year} {2014})}\BibitemShut {NoStop}%
\bibitem [{\citenamefont {Lansbergen}\ \emph {et~al.}(2012)\citenamefont
  {Lansbergen}, \citenamefont {Ono},\ and\ \citenamefont
  {Fujiwara}}]{lansbergen2012donor}%
  \BibitemOpen
  \bibfield  {author} {\bibinfo {author} {\bibfnamefont {G.P.}\ \bibnamefont
  {Lansbergen}}, \bibinfo {author} {\bibfnamefont {Y.}~\bibnamefont {Ono}}, \
  and\ \bibinfo {author} {\bibfnamefont {A.}~\bibnamefont {Fujiwara}},\
  }\bibfield  {title} {\enquote {\bibinfo {title} {Donor-based single electron
  pumps with tunable donor binding energy},}\ }\href@noop {} {\bibfield
  {journal} {\bibinfo  {journal} {Nano Lett.}\ }\textbf {\bibinfo {volume}
  {12}},\ \bibinfo {pages} {763--768} (\bibinfo {year} {2012})}\BibitemShut
  {NoStop}%
\bibitem [{\citenamefont {Roche}\ \emph {et~al.}(2013)\citenamefont {Roche},
  \citenamefont {Riwar}, \citenamefont {Voisin}, \citenamefont
  {Dupont-Ferrier}, \citenamefont {Wacquez}, \citenamefont {Vinet},
  \citenamefont {Sanquer}, \citenamefont {Splettstoesser},\ and\ \citenamefont
  {Jehl}}]{roche2013two}%
  \BibitemOpen
  \bibfield  {author} {\bibinfo {author} {\bibfnamefont {B.}~\bibnamefont
  {Roche}}, \bibinfo {author} {\bibfnamefont {R-P.}\ \bibnamefont {Riwar}},
  \bibinfo {author} {\bibfnamefont {B.}~\bibnamefont {Voisin}}, \bibinfo
  {author} {\bibfnamefont {E.}~\bibnamefont {Dupont-Ferrier}}, \bibinfo
  {author} {\bibfnamefont {R.}~\bibnamefont {Wacquez}}, \bibinfo {author}
  {\bibfnamefont {M.}~\bibnamefont {Vinet}}, \bibinfo {author} {\bibfnamefont
  {M.}~\bibnamefont {Sanquer}}, \bibinfo {author} {\bibfnamefont
  {J.}~\bibnamefont {Splettstoesser}}, \ and\ \bibinfo {author} {\bibfnamefont
  {X.}~\bibnamefont {Jehl}},\ }\bibfield  {title} {\enquote {\bibinfo {title}
  {A two-atom electron pump},}\ }\href@noop {} {\bibfield  {journal} {\bibinfo
  {journal} {Nat Commun.}\ }\textbf {\bibinfo {volume} {4}},\ \bibinfo {pages}
  {1581} (\bibinfo {year} {2013})}\BibitemShut {NoStop}%
\bibitem [{\citenamefont {Bae}\ \emph {et~al.}(2015)\citenamefont {Bae},
  \citenamefont {Ahn}, \citenamefont {Seo}, \citenamefont {Chung},
  \citenamefont {Fletcher}, \citenamefont {Giblin}, \citenamefont {Kataoka},\
  and\ \citenamefont {Kim}}]{KRISS2015}%
  \BibitemOpen
  \bibfield  {author} {\bibinfo {author} {\bibfnamefont {Myung-Ho}\
  \bibnamefont {Bae}}, \bibinfo {author} {\bibfnamefont {Ye-Hwan}\ \bibnamefont
  {Ahn}}, \bibinfo {author} {\bibfnamefont {Minky}\ \bibnamefont {Seo}},
  \bibinfo {author} {\bibfnamefont {Yunchul}\ \bibnamefont {Chung}}, \bibinfo
  {author} {\bibfnamefont {J~D}\ \bibnamefont {Fletcher}}, \bibinfo {author}
  {\bibfnamefont {S~P}\ \bibnamefont {Giblin}}, \bibinfo {author}
  {\bibfnamefont {M}~\bibnamefont {Kataoka}}, \ and\ \bibinfo {author}
  {\bibfnamefont {Nam}\ \bibnamefont {Kim}},\ }\bibfield  {title} {\enquote
  {\bibinfo {title} {Precision measurement of a potential-profile tunable
  single-electron pump},}\ }\href@noop {} {\bibfield  {journal} {\bibinfo
  {journal} {Metrologia}\ }\textbf {\bibinfo {volume} {52}},\ \bibinfo {pages}
  {195} (\bibinfo {year} {2015})}\BibitemShut {NoStop}%
\bibitem [{\citenamefont {Stein}\ \emph {et~al.}(2016)\citenamefont {Stein},
  \citenamefont {Scherer}, \citenamefont {Gerster}, \citenamefont {Behr},
  \citenamefont {G{\"o}tz}, \citenamefont {Pesel}, \citenamefont {Leicht},
  \citenamefont {Ubbelohde}, \citenamefont {Weimann}, \citenamefont {Pierz}
  \emph {et~al.}}]{stein2016robustness}%
  \BibitemOpen
  \bibfield  {author} {\bibinfo {author} {\bibfnamefont {F}~\bibnamefont
  {Stein}}, \bibinfo {author} {\bibfnamefont {H}~\bibnamefont {Scherer}},
  \bibinfo {author} {\bibfnamefont {T}~\bibnamefont {Gerster}}, \bibinfo
  {author} {\bibfnamefont {R}~\bibnamefont {Behr}}, \bibinfo {author}
  {\bibfnamefont {M}~\bibnamefont {G{\"o}tz}}, \bibinfo {author} {\bibfnamefont
  {E}~\bibnamefont {Pesel}}, \bibinfo {author} {\bibfnamefont {C}~\bibnamefont
  {Leicht}}, \bibinfo {author} {\bibfnamefont {N}~\bibnamefont {Ubbelohde}},
  \bibinfo {author} {\bibfnamefont {T}~\bibnamefont {Weimann}}, \bibinfo
  {author} {\bibfnamefont {K}~\bibnamefont {Pierz}},  \emph {et~al.},\
  }\bibfield  {title} {\enquote {\bibinfo {title} {Robustness of
  single-electron pumps at sub-ppm current accuracy level},}\ }\href@noop {}
  {\bibfield  {journal} {\bibinfo  {journal} {Metrologia}\ }\textbf {\bibinfo
  {volume} {54}},\ \bibinfo {pages} {S1} (\bibinfo {year} {2016})}\BibitemShut
  {NoStop}%
\bibitem [{\citenamefont {Giblin}\ \emph {et~al.}(2017)\citenamefont {Giblin},
  \citenamefont {Bae}, \citenamefont {Kim}, \citenamefont {Ahn},\ and\
  \citenamefont {Kataoka}}]{giblin2017robust}%
  \BibitemOpen
  \bibfield  {author} {\bibinfo {author} {\bibfnamefont {S.P.}\ \bibnamefont
  {Giblin}}, \bibinfo {author} {\bibfnamefont {M.H.}\ \bibnamefont {Bae}},
  \bibinfo {author} {\bibfnamefont {N.}~\bibnamefont {Kim}}, \bibinfo {author}
  {\bibfnamefont {Ye-Hwan}\ \bibnamefont {Ahn}}, \ and\ \bibinfo {author}
  {\bibfnamefont {M.}~\bibnamefont {Kataoka}},\ }\bibfield  {title} {\enquote
  {\bibinfo {title} {Robust operation of a gaas tunable barrier electron
  pump},}\ }\href@noop {} {\bibfield  {journal} {\bibinfo  {journal}
  {Metrologia}\ }\textbf {\bibinfo {volume} {54}},\ \bibinfo {pages} {299}
  (\bibinfo {year} {2017})}\BibitemShut {NoStop}%
\bibitem [{\citenamefont {Kashcheyevs}\ and\ \citenamefont
  {Kaestner}(2010)}]{decaycascade}%
  \BibitemOpen
  \bibfield  {author} {\bibinfo {author} {\bibfnamefont {Vyacheslavs}\
  \bibnamefont {Kashcheyevs}}\ and\ \bibinfo {author} {\bibfnamefont {Bernd}\
  \bibnamefont {Kaestner}},\ }\bibfield  {title} {\enquote {\bibinfo {title}
  {Universal decay cascade model for dynamic quantum dot initialization},}\
  }\href {\doibase 10.1103/PhysRevLett.104.186805} {\bibfield  {journal}
  {\bibinfo  {journal} {Phys. Rev. Lett.}\ }\textbf {\bibinfo {volume} {104}},\
  \bibinfo {pages} {186805} (\bibinfo {year} {2010})}\BibitemShut {NoStop}%
\bibitem [{\citenamefont {Yamahata}\ \emph
  {et~al.}(2014{\natexlab{b}})\citenamefont {Yamahata}, \citenamefont
  {Nishiguchi},\ and\ \citenamefont {Fujiwara}}]{GentoCrossOver}%
  \BibitemOpen
  \bibfield  {author} {\bibinfo {author} {\bibfnamefont {Gento}\ \bibnamefont
  {Yamahata}}, \bibinfo {author} {\bibfnamefont {Katsuhiko}\ \bibnamefont
  {Nishiguchi}}, \ and\ \bibinfo {author} {\bibfnamefont {Akira}\ \bibnamefont
  {Fujiwara}},\ }\bibfield  {title} {\enquote {\bibinfo {title} {Accuracy
  evaluation and mechanism crossover of single-electron transfer in si
  tunable-barrier turnstiles},}\ }\href@noop {} {\bibfield  {journal} {\bibinfo
   {journal} {Phys. Rev. B}\ }\textbf {\bibinfo {volume} {89}},\ \bibinfo
  {pages} {165302} (\bibinfo {year} {2014}{\natexlab{b}})}\BibitemShut
  {NoStop}%
\bibitem [{\citenamefont {Kashcheyevs}\ and\ \citenamefont
  {Timoshenko}(2014)}]{kashcheyevs2014modeling}%
  \BibitemOpen
  \bibfield  {author} {\bibinfo {author} {\bibfnamefont {Vyacheslavs}\
  \bibnamefont {Kashcheyevs}}\ and\ \bibinfo {author} {\bibfnamefont {Janis}\
  \bibnamefont {Timoshenko}},\ }\bibfield  {title} {\enquote {\bibinfo {title}
  {Modeling of a tunable-barrier non-adiabatic electron pump beyond the decay
  cascade model},}\ }in\ \href@noop {} {\emph {\bibinfo {booktitle} {Precision
  Electromagnetic Measurements (CPEM 2014), 2014 Conference on}}}\ (\bibinfo
  {organization} {IEEE},\ \bibinfo {year} {2014})\ pp.\ \bibinfo {pages}
  {536--537}\BibitemShut {NoStop}%
\bibitem [{\citenamefont {Fricke}\ \emph {et~al.}(2013)\citenamefont {Fricke},
  \citenamefont {Wulf}, \citenamefont {Kaestner}, \citenamefont {Kashcheyevs},
  \citenamefont {Timoshenko}, \citenamefont {Nazarov}, \citenamefont {Hohls},
  \citenamefont {Mirovsky}, \citenamefont {Mackrodt}, \citenamefont {Dolata},
  \citenamefont {Weimann}, \citenamefont {Pierz},\ and\ \citenamefont
  {Schumacher}}]{lukasCountingStat}%
  \BibitemOpen
  \bibfield  {author} {\bibinfo {author} {\bibfnamefont {Lukas}\ \bibnamefont
  {Fricke}}, \bibinfo {author} {\bibfnamefont {Michael}\ \bibnamefont {Wulf}},
  \bibinfo {author} {\bibfnamefont {Bernd}\ \bibnamefont {Kaestner}}, \bibinfo
  {author} {\bibfnamefont {Vyacheslavs}\ \bibnamefont {Kashcheyevs}}, \bibinfo
  {author} {\bibfnamefont {Janis}\ \bibnamefont {Timoshenko}}, \bibinfo
  {author} {\bibfnamefont {Pavel}\ \bibnamefont {Nazarov}}, \bibinfo {author}
  {\bibfnamefont {Frank}\ \bibnamefont {Hohls}}, \bibinfo {author}
  {\bibfnamefont {Philipp}\ \bibnamefont {Mirovsky}}, \bibinfo {author}
  {\bibfnamefont {Brigitte}\ \bibnamefont {Mackrodt}}, \bibinfo {author}
  {\bibfnamefont {Ralf}\ \bibnamefont {Dolata}}, \bibinfo {author}
  {\bibfnamefont {Thomas}\ \bibnamefont {Weimann}}, \bibinfo {author}
  {\bibfnamefont {Klaus}\ \bibnamefont {Pierz}}, \ and\ \bibinfo {author}
  {\bibfnamefont {Hans~W.}\ \bibnamefont {Schumacher}},\ }\bibfield  {title}
  {\enquote {\bibinfo {title} {Counting statistics for electron capture in a
  dynamic quantum dot},}\ }\href@noop {} {\bibfield  {journal} {\bibinfo
  {journal} {Phys. Rev. Lett.}\ }\textbf {\bibinfo {volume} {110}},\ \bibinfo
  {pages} {126803} (\bibinfo {year} {2013})}\BibitemShut {NoStop}%
\bibitem [{\citenamefont {Angus}\ \emph {et~al.}(2007)\citenamefont {Angus},
  \citenamefont {Ferguson}, \citenamefont {Dzurak},\ and\ \citenamefont
  {Clark}}]{angus2007gate}%
  \BibitemOpen
  \bibfield  {author} {\bibinfo {author} {\bibfnamefont {Susan~J}\ \bibnamefont
  {Angus}}, \bibinfo {author} {\bibfnamefont {Andrew~J}\ \bibnamefont
  {Ferguson}}, \bibinfo {author} {\bibfnamefont {Andrew~S}\ \bibnamefont
  {Dzurak}}, \ and\ \bibinfo {author} {\bibfnamefont {Robert~G}\ \bibnamefont
  {Clark}},\ }\bibfield  {title} {\enquote {\bibinfo {title} {Gate-defined
  quantum dots in intrinsic silicon},}\ }\href@noop {} {\bibfield  {journal}
  {\bibinfo  {journal} {Nano Lett.}\ }\textbf {\bibinfo {volume} {7}},\
  \bibinfo {pages} {2051--2055} (\bibinfo {year} {2007})}\BibitemShut {NoStop}%
\bibitem [{\citenamefont {Rossi}\ \emph {et~al.}(2015)\citenamefont {Rossi},
  \citenamefont {Tanttu}, \citenamefont {Hudson}, \citenamefont {Sun},
  \citenamefont {M{\"o}tt{\"o}nen},\ and\ \citenamefont
  {Dzurak}}]{rossi2015silicon}%
  \BibitemOpen
  \bibfield  {author} {\bibinfo {author} {\bibfnamefont {Alessandro}\
  \bibnamefont {Rossi}}, \bibinfo {author} {\bibfnamefont {Tuomo}\ \bibnamefont
  {Tanttu}}, \bibinfo {author} {\bibfnamefont {Fay~E.}\ \bibnamefont {Hudson}},
  \bibinfo {author} {\bibfnamefont {Yuxin}\ \bibnamefont {Sun}}, \bibinfo
  {author} {\bibfnamefont {Mikko}\ \bibnamefont {M{\"o}tt{\"o}nen}}, \ and\
  \bibinfo {author} {\bibfnamefont {Andrew~S.}\ \bibnamefont {Dzurak}},\
  }\bibfield  {title} {\enquote {\bibinfo {title} {Silicon
  metal-oxide-semiconductor quantum dots for single-electron pumping},}\
  }\href@noop {} {\bibfield  {journal} {\bibinfo  {journal} {J. Vis. Exp.}\
  }\textbf {\bibinfo {volume} {100}},\ \bibinfo {pages} {e52852} (\bibinfo
  {year} {2015})}\BibitemShut {NoStop}%
\bibitem [{\citenamefont {Law}\ and\ \citenamefont {Cea}(1998)}]{tcad}%
  \BibitemOpen
  \bibfield  {author} {\bibinfo {author} {\bibfnamefont {Mark~E}\ \bibnamefont
  {Law}}\ and\ \bibinfo {author} {\bibfnamefont {Stephen~M}\ \bibnamefont
  {Cea}},\ }\bibfield  {title} {\enquote {\bibinfo {title} {Continuum based
  modeling of silicon integrated circuit processing: An object oriented
  approach},}\ }\href@noop {} {\bibfield  {journal} {\bibinfo  {journal}
  {Computational Materials Science}\ }\textbf {\bibinfo {volume} {12}},\
  \bibinfo {pages} {289--308} (\bibinfo {year} {1998})}\BibitemShut {NoStop}%
\bibitem [{\citenamefont {Pothier}\ \emph {et~al.}(1992)\citenamefont
  {Pothier}, \citenamefont {Lafarge}, \citenamefont {Urbina}, \citenamefont
  {Esteve},\ and\ \citenamefont {Devoret}}]{pothier1992single}%
  \BibitemOpen
  \bibfield  {author} {\bibinfo {author} {\bibfnamefont {H.}~\bibnamefont
  {Pothier}}, \bibinfo {author} {\bibfnamefont {P.}~\bibnamefont {Lafarge}},
  \bibinfo {author} {\bibfnamefont {C.}~\bibnamefont {Urbina}}, \bibinfo
  {author} {\bibfnamefont {D.}~\bibnamefont {Esteve}}, \ and\ \bibinfo {author}
  {\bibfnamefont {M.H.}\ \bibnamefont {Devoret}},\ }\bibfield  {title}
  {\enquote {\bibinfo {title} {Single-electron pump based on charging
  effects},}\ }\href@noop {} {\bibfield  {journal} {\bibinfo  {journal} {EPL}\
  }\textbf {\bibinfo {volume} {17}},\ \bibinfo {pages} {249} (\bibinfo {year}
  {1992})}\BibitemShut {NoStop}%
\bibitem [{sup()}]{suppinfo}%
  \BibitemOpen
  \href@noop {} {\ }\bibinfo {note} {See Supplemental Material for more
  information about the addition energy extraction, TCAD simulation as well as
  the precision measurement technique.}\BibitemShut {Stop}%
\bibitem [{\citenamefont {Kataoka}\ \emph {et~al.}(2011)\citenamefont
  {Kataoka}, \citenamefont {Fletcher}, \citenamefont {See}, \citenamefont
  {Giblin}, \citenamefont {Janssen}, \citenamefont {Griffiths}, \citenamefont
  {Jones}, \citenamefont {Farrer},\ and\ \citenamefont
  {Ritchie}}]{Masaya2011PRL}%
  \BibitemOpen
  \bibfield  {author} {\bibinfo {author} {\bibfnamefont {M.}~\bibnamefont
  {Kataoka}}, \bibinfo {author} {\bibfnamefont {J.~D.}\ \bibnamefont
  {Fletcher}}, \bibinfo {author} {\bibfnamefont {P.}~\bibnamefont {See}},
  \bibinfo {author} {\bibfnamefont {S.~P.}\ \bibnamefont {Giblin}}, \bibinfo
  {author} {\bibfnamefont {T.~J. B.~M.}\ \bibnamefont {Janssen}}, \bibinfo
  {author} {\bibfnamefont {J.~P.}\ \bibnamefont {Griffiths}}, \bibinfo {author}
  {\bibfnamefont {G.~A.~C.}\ \bibnamefont {Jones}}, \bibinfo {author}
  {\bibfnamefont {I.}~\bibnamefont {Farrer}}, \ and\ \bibinfo {author}
  {\bibfnamefont {D.~A.}\ \bibnamefont {Ritchie}},\ }\bibfield  {title}
  {\enquote {\bibinfo {title} {Tunable nonadiabatic excitation in a
  single-electron quantum dot},}\ }\href {\doibase
  10.1103/PhysRevLett.106.126801} {\bibfield  {journal} {\bibinfo  {journal}
  {Phys. Rev. Lett.}\ }\textbf {\bibinfo {volume} {106}},\ \bibinfo {pages}
  {126801} (\bibinfo {year} {2011})}\BibitemShut {NoStop}%
\bibitem [{\citenamefont {Fletcher}\ \emph {et~al.}(2012)\citenamefont
  {Fletcher}, \citenamefont {Kataoka}, \citenamefont {Giblin}, \citenamefont
  {Park}, \citenamefont {Sim}, \citenamefont {See}, \citenamefont {Ritchie},
  \citenamefont {Griffiths}, \citenamefont {Jones}, \citenamefont {Beere},\
  and\ \citenamefont {Janssen}}]{Fletcher2012PRB}%
  \BibitemOpen
  \bibfield  {author} {\bibinfo {author} {\bibfnamefont {J.~D.}\ \bibnamefont
  {Fletcher}}, \bibinfo {author} {\bibfnamefont {M.}~\bibnamefont {Kataoka}},
  \bibinfo {author} {\bibfnamefont {S.~P.}\ \bibnamefont {Giblin}}, \bibinfo
  {author} {\bibfnamefont {Sunghun}\ \bibnamefont {Park}}, \bibinfo {author}
  {\bibfnamefont {H.-S.}\ \bibnamefont {Sim}}, \bibinfo {author} {\bibfnamefont
  {P.}~\bibnamefont {See}}, \bibinfo {author} {\bibfnamefont {D.~A.}\
  \bibnamefont {Ritchie}}, \bibinfo {author} {\bibfnamefont {J.~P.}\
  \bibnamefont {Griffiths}}, \bibinfo {author} {\bibfnamefont {G.~A.~C.}\
  \bibnamefont {Jones}}, \bibinfo {author} {\bibfnamefont {H.~E.}\ \bibnamefont
  {Beere}}, \ and\ \bibinfo {author} {\bibfnamefont {T.~J. B.~M.}\ \bibnamefont
  {Janssen}},\ }\bibfield  {title} {\enquote {\bibinfo {title} {Stabilization
  of single-electron pumps by high magnetic fields},}\ }\href {\doibase
  10.1103/PhysRevB.86.155311} {\bibfield  {journal} {\bibinfo  {journal} {Phys.
  Rev. B}\ }\textbf {\bibinfo {volume} {86}},\ \bibinfo {pages} {155311}
  (\bibinfo {year} {2012})}\BibitemShut {NoStop}%
\bibitem [{\citenamefont {Giblin}()}]{Giblin201xCCC}%
  \BibitemOpen
  \bibfield  {author} {\bibinfo {author} {\bibfnamefont {SP}~\bibnamefont
  {Giblin}},\ }\href@noop {} {\ }\bibinfo {note} {Manuscript in
  preparation}\BibitemShut {NoStop}%
\bibitem [{\citenamefont {Drung}\ \emph {et~al.}(2015)\citenamefont {Drung},
  \citenamefont {Krause}, \citenamefont {Giblin}, \citenamefont {Djordjevic},
  \citenamefont {Piquemal}, \citenamefont {S{\'e}ron}, \citenamefont {Rengnez},
  \citenamefont {G{\"o}tz}, \citenamefont {Pesel},\ and\ \citenamefont
  {Scherer}}]{drung2015validation}%
  \BibitemOpen
  \bibfield  {author} {\bibinfo {author} {\bibfnamefont {Dietmar}\ \bibnamefont
  {Drung}}, \bibinfo {author} {\bibfnamefont {Christian}\ \bibnamefont
  {Krause}}, \bibinfo {author} {\bibfnamefont {Stephen~P}\ \bibnamefont
  {Giblin}}, \bibinfo {author} {\bibfnamefont {Sophie}\ \bibnamefont
  {Djordjevic}}, \bibinfo {author} {\bibfnamefont {Francois}\ \bibnamefont
  {Piquemal}}, \bibinfo {author} {\bibfnamefont {Olivier}\ \bibnamefont
  {S{\'e}ron}}, \bibinfo {author} {\bibfnamefont {Florentin}\ \bibnamefont
  {Rengnez}}, \bibinfo {author} {\bibfnamefont {Martin}\ \bibnamefont
  {G{\"o}tz}}, \bibinfo {author} {\bibfnamefont {Eckart}\ \bibnamefont
  {Pesel}}, \ and\ \bibinfo {author} {\bibfnamefont {Hansj{\"o}rg}\
  \bibnamefont {Scherer}},\ }\bibfield  {title} {\enquote {\bibinfo {title}
  {Validation of the ultrastable low-noise current amplifier as travelling
  standard for small direct currents},}\ }\href@noop {} {\bibfield  {journal}
  {\bibinfo  {journal} {Metrologia}\ }\textbf {\bibinfo {volume} {52}},\
  \bibinfo {pages} {756} (\bibinfo {year} {2015})}\BibitemShut {NoStop}%
\bibitem [{\citenamefont {Zimmerman}(1998)}]{zimmerman1998primer}%
  \BibitemOpen
  \bibfield  {author} {\bibinfo {author} {\bibfnamefont {N.~M.}\ \bibnamefont
  {Zimmerman}},\ }\bibfield  {title} {\enquote {\bibinfo {title} {A primer on
  electrical units in the systeme international},}\ }\href@noop {} {\bibfield
  {journal} {\bibinfo  {journal} {American Journal of Physics}\ }\textbf
  {\bibinfo {volume} {66}},\ \bibinfo {pages} {324} (\bibinfo {year}
  {1998})}\BibitemShut {NoStop}%
\bibitem [{\citenamefont {Mohr}\ \emph {et~al.}(2016)\citenamefont {Mohr},
  \citenamefont {Newell},\ and\ \citenamefont {Taylor}}]{mohr2016codata}%
  \BibitemOpen
  \bibfield  {author} {\bibinfo {author} {\bibfnamefont {Peter~J}\ \bibnamefont
  {Mohr}}, \bibinfo {author} {\bibfnamefont {David~B}\ \bibnamefont {Newell}},
  \ and\ \bibinfo {author} {\bibfnamefont {Barry~N}\ \bibnamefont {Taylor}},\
  }\bibfield  {title} {\enquote {\bibinfo {title} {Codata recommended values of
  the fundamental physical constants: 2014},}\ }\href@noop {} {\bibfield
  {journal} {\bibinfo  {journal} {Journal of Physical and Chemical Reference
  Data}\ }\textbf {\bibinfo {volume} {45}},\ \bibinfo {pages} {043102}
  (\bibinfo {year} {2016})},\ \bibinfo {note} {reference on page
  65}\BibitemShut {NoStop}%
\end{thebibliography}%


\begin{thebibliography}{5}%
\makeatletter
\providecommand \@ifxundefined [1]{%
 \@ifx{#1\undefined}
}%
\providecommand \@ifnum [1]{%
 \ifnum #1\expandafter \@firstoftwo
 \else \expandafter \@secondoftwo
 \fi
}%
\providecommand \@ifx [1]{%
 \ifx #1\expandafter \@firstoftwo
 \else \expandafter \@secondoftwo
 \fi
}%
\providecommand \natexlab [1]{#1}%
\providecommand \enquote  [1]{``#1''}%
\providecommand \bibnamefont  [1]{#1}%
\providecommand \bibfnamefont [1]{#1}%
\providecommand \citenamefont [1]{#1}%
\providecommand \href@noop [0]{\@secondoftwo}%
\providecommand \href [0]{\begingroup \@sanitize@url \@href}%
\providecommand \@href[1]{\@@startlink{#1}\@@href}%
\providecommand \@@href[1]{\endgroup#1\@@endlink}%
\providecommand \@sanitize@url [0]{\catcode `\\12\catcode `\$12\catcode
  `\&12\catcode `\#12\catcode `\^12\catcode `\_12\catcode `\%12\relax}%
\providecommand \@@startlink[1]{}%
\providecommand \@@endlink[0]{}%
\providecommand \url  [0]{\begingroup\@sanitize@url \@url }%
\providecommand \@url [1]{\endgroup\@href {#1}{\urlprefix }}%
\providecommand \urlprefix  [0]{URL }%
\providecommand \Eprint [0]{\href }%
\providecommand \doibase [0]{http://dx.doi.org/}%
\providecommand \selectlanguage [0]{\@gobble}%
\providecommand \bibinfo  [0]{\@secondoftwo}%
\providecommand \bibfield  [0]{\@secondoftwo}%
\providecommand \translation [1]{[#1]}%
\providecommand \BibitemOpen [0]{}%
\providecommand \bibitemStop [0]{}%
\providecommand \bibitemNoStop [0]{.\EOS\space}%
\providecommand \EOS [0]{\spacefactor3000\relax}%
\providecommand \BibitemShut  [1]{\csname bibitem#1\endcsname}%
\let\auto@bib@innerbib\@empty
\bibitem [{\citenamefont {Law}\ and\ \citenamefont {Cea}(1998)}]{tcad}%
  \BibitemOpen
  \bibfield  {author} {\bibinfo {author} {\bibfnamefont {Mark~E}\ \bibnamefont
  {Law}}\ and\ \bibinfo {author} {\bibfnamefont {Stephen~M}\ \bibnamefont
  {Cea}},\ }\bibfield  {title} {\enquote {\bibinfo {title} {Continuum based
  modeling of silicon integrated circuit processing: An object oriented
  approach},}\ }\href@noop {} {\bibfield  {journal} {\bibinfo  {journal}
  {Computational Materials Science}\ }\textbf {\bibinfo {volume} {12}},\
  \bibinfo {pages} {289--308} (\bibinfo {year} {1998})}\BibitemShut {NoStop}%
\bibitem [{\citenamefont {Thorbeck}\ and\ \citenamefont
  {Zimmerman}(2015)}]{thorbeck2015formation}%
  \BibitemOpen
  \bibfield  {author} {\bibinfo {author} {\bibfnamefont {Ted}\ \bibnamefont
  {Thorbeck}}\ and\ \bibinfo {author} {\bibfnamefont {Neil~M}\ \bibnamefont
  {Zimmerman}},\ }\bibfield  {title} {\enquote {\bibinfo {title} {Formation of
  strain-induced quantum dots in gated semiconductor nanostructures},}\
  }\href@noop {} {\bibfield  {journal} {\bibinfo  {journal} {AIP Advances}\
  }\textbf {\bibinfo {volume} {5}},\ \bibinfo {pages} {087107} (\bibinfo {year}
  {2015})}\BibitemShut {NoStop}%
\bibitem [{\citenamefont {Rossi}\ \emph {et~al.}(2014)\citenamefont {Rossi},
  \citenamefont {Tanttu}, \citenamefont {Tan}, \citenamefont {Iisakka},
  \citenamefont {Zhao}, \citenamefont {Chan}, \citenamefont {Tettamanzi},
  \citenamefont {Rogge}, \citenamefont {Dzurak},\ and\ \citenamefont
  {M\"ott\"onen}}]{Ale2014}%
  \BibitemOpen
  \bibfield  {author} {\bibinfo {author} {\bibfnamefont {Alessandro}\
  \bibnamefont {Rossi}}, \bibinfo {author} {\bibfnamefont {Tuomo}\ \bibnamefont
  {Tanttu}}, \bibinfo {author} {\bibfnamefont {Kuan~Yen}\ \bibnamefont {Tan}},
  \bibinfo {author} {\bibfnamefont {Ilkka}\ \bibnamefont {Iisakka}}, \bibinfo
  {author} {\bibfnamefont {Ruichen}\ \bibnamefont {Zhao}}, \bibinfo {author}
  {\bibfnamefont {Kok~Wai}\ \bibnamefont {Chan}}, \bibinfo {author}
  {\bibfnamefont {Giuseppe~C}\ \bibnamefont {Tettamanzi}}, \bibinfo {author}
  {\bibfnamefont {Sven}\ \bibnamefont {Rogge}}, \bibinfo {author}
  {\bibfnamefont {Andrew~S}\ \bibnamefont {Dzurak}}, \ and\ \bibinfo {author}
  {\bibfnamefont {Mikko}\ \bibnamefont {M\"ott\"onen}},\ }\bibfield  {title}
  {\enquote {\bibinfo {title} {An accurate single-electron pump based on a
  highly tunable silicon quantum dot},}\ }\href@noop {} {\bibfield  {journal}
  {\bibinfo  {journal} {Nano Lett.}\ }\textbf {\bibinfo {volume} {14}},\
  \bibinfo {pages} {3405--3411} (\bibinfo {year} {2014})}\BibitemShut {NoStop}%
\bibitem [{\citenamefont {Allan}(1987)}]{allan1987should}%
  \BibitemOpen
  \bibfield  {author} {\bibinfo {author} {\bibfnamefont {David~W}\ \bibnamefont
  {Allan}},\ }\bibfield  {title} {\enquote {\bibinfo {title} {Should the
  classical variance be used as a basic measure in standards metrology?}}\
  }\href@noop {} {\bibfield  {journal} {\bibinfo  {journal} {IEEE Transactions
  on instrumentation and measurement}\ }\textbf {\bibinfo {volume} {1001}},\
  \bibinfo {pages} {646--654} (\bibinfo {year} {1987})}\BibitemShut {NoStop}%
\bibitem [{\citenamefont {Giblin}\ \emph {et~al.}(2017)\citenamefont {Giblin},
  \citenamefont {Bae}, \citenamefont {Kim}, \citenamefont {Ahn},\ and\
  \citenamefont {Kataoka}}]{giblin2017robust}%
  \BibitemOpen
  \bibfield  {author} {\bibinfo {author} {\bibfnamefont {SP}~\bibnamefont
  {Giblin}}, \bibinfo {author} {\bibfnamefont {MH}~\bibnamefont {Bae}},
  \bibinfo {author} {\bibfnamefont {N}~\bibnamefont {Kim}}, \bibinfo {author}
  {\bibfnamefont {Ye-Hwan}\ \bibnamefont {Ahn}}, \ and\ \bibinfo {author}
  {\bibfnamefont {M}~\bibnamefont {Kataoka}},\ }\bibfield  {title} {\enquote
  {\bibinfo {title} {Robust operation of a gaas tunable barrier electron
  pump},}\ }\href@noop {} {\bibfield  {journal} {\bibinfo  {journal}
  {Metrologia}\ }\textbf {\bibinfo {volume} {54}},\ \bibinfo {pages} {299}
  (\bibinfo {year} {2017})}\BibitemShut {NoStop}%
\end{thebibliography}%
\onecolumngrid
\pagebreak

\end{document}



\title{Supplementary Information: Thermal-error regime in high-accuracy gigahertz single-electron pumping}



\author{R.~Zhao}
\email[]{ruichen.zhao@student.unsw.edu.au}
\affiliation{School of Electrical Engineering and Telecommunications, University of New South Wales, Sydney, New South Wales 2052, Australia}

\author{A.~Rossi}
\affiliation{Cavendish Laboratory, University of Cambridge, J J Thomson Avenue, Cambridge CB3 0HE, United Kingdom}

\author{S.~P.~Giblin}
\affiliation{National Physical Laboratory, Hampton Road, Teddington, Middlesex TW11 0LW, United Kingdom}

\author{J.~D.~Fletcher}
\affiliation{National Physical Laboratory, Hampton Road, Teddington, Middlesex TW11 0LW, United Kingdom}

\author{F.~E.~Hudson}
\affiliation{School of Electrical Engineering and Telecommunications, University of New South Wales, Sydney, New South Wales 2052, Australia}

\author{M.~M\"ott\"onen}
\affiliation{QCD Labs, COMP Centre of Excellence, Department of Applied Physics, Aalto University, 00076 AALTO, Finland}

\author{M.~Kataoka}
\affiliation{National Physical Laboratory, Hampton Road, Teddington, Middlesex TW11 0LW, United Kingdom}

\author{A.~S.~Dzurak}
\affiliation{School of Electrical Engineering and Telecommunications, University of New South Wales, Sydney, New South Wales 2052, Australia}


\date{\today}
\maketitle

\onecolumngrid
\pagebreak

 \setcounter{figure}{0}   
 \makeatletter 
 \renewcommand{\thefigure}{S\@arabic\c@figure}
 \makeatother
 \setcounter{equation}{0}   
 \makeatletter 
 \renewcommand{\theequation}{S\@arabic\c@equation}
 \makeatother

\section{Conduction Band Energy Profile Simulation}
In this work, we estimate the addition energy of the QD based on the conduction band energy profile ($E_\textmd{CB}$) simulated using the commercial semiconductor software ISE-TCAD \cite{tcad}. This package can replicate the electrostatic potential profile, in bulk semiconductors, by iteratively solving Poisson's equation attached to the user-defined mesh grid at low temperatures. The software package can also conveniently translate the electrostatic potential profile into the $E_\textmd{CB}$ landscape. The simulated model, based on the geometry of the device used in the experiments, is presented in Fig.~\ref{fig:tcad}(a). The measured threshold voltage of $0.4$ V corresponds to a Si/SiO$_2$ interface charge density of $Q_\textmd{ox} = -1.45\times10^{11}$ cm$^{-2}$ in the TCAD simulation. The metal-gate-induced strain, not considered by the ISE-TCAD package, could potentially affect the QD formation in our device architecture \cite{thorbeck2015formation}. For the purpose of matching the simulation to the experiment, we choose to compensate the effect of the strain through adjustment in gate voltages. Note that our previous work clearly demonstrates that the enhancement of the addition energy imposed by the gate C1 and C2 is electrostatic \cite{Ale2014}. Hence, we assume that the enhancement in electrostatic confinement is not affected by strain effects, and can be accurately captured in a pure-electrostatic simulation using the TCAD.  We calibrate the gate voltages as follows: First, we simulate the $E_\textmd{CB}$ profile of a QD similar to the one measured in a dc source-drain bias scan in Fig.~\ref{fig:tcad}(b). The measurement is carried out in the weak electrostatic confinement regime, where the source-drain conductance is still measurable. We use all gate voltages from the experiment and keep the source-drain bias voltage, $V_\textmd{DS}$, at zero. We adjust the voltages $V_\textmd{B1}$ and $V_\textmd{B2}$ in the simulation until the estimated addition energy of the QD matches with the experimental result ($\approx 3$ meV). Using these calibrated barrier gate voltages ($V_\textmd{B1} = V_\textmd{B2} = 0.34$ V), we simulated the $E_\textmd{CB}$ profile of the QD under the condition that closely resembles the SE pumping experiment in Fig.~\ref{fig:Plateaux}. The sliced $E_\textmd{CB}$ profiles through the center of the QD are presented in Fig.~\ref{fig:tcad}(c) and (d).

\section{Estimation of the Addition Energy}

\begin{figure}
	\includegraphics[width=0.9\linewidth]{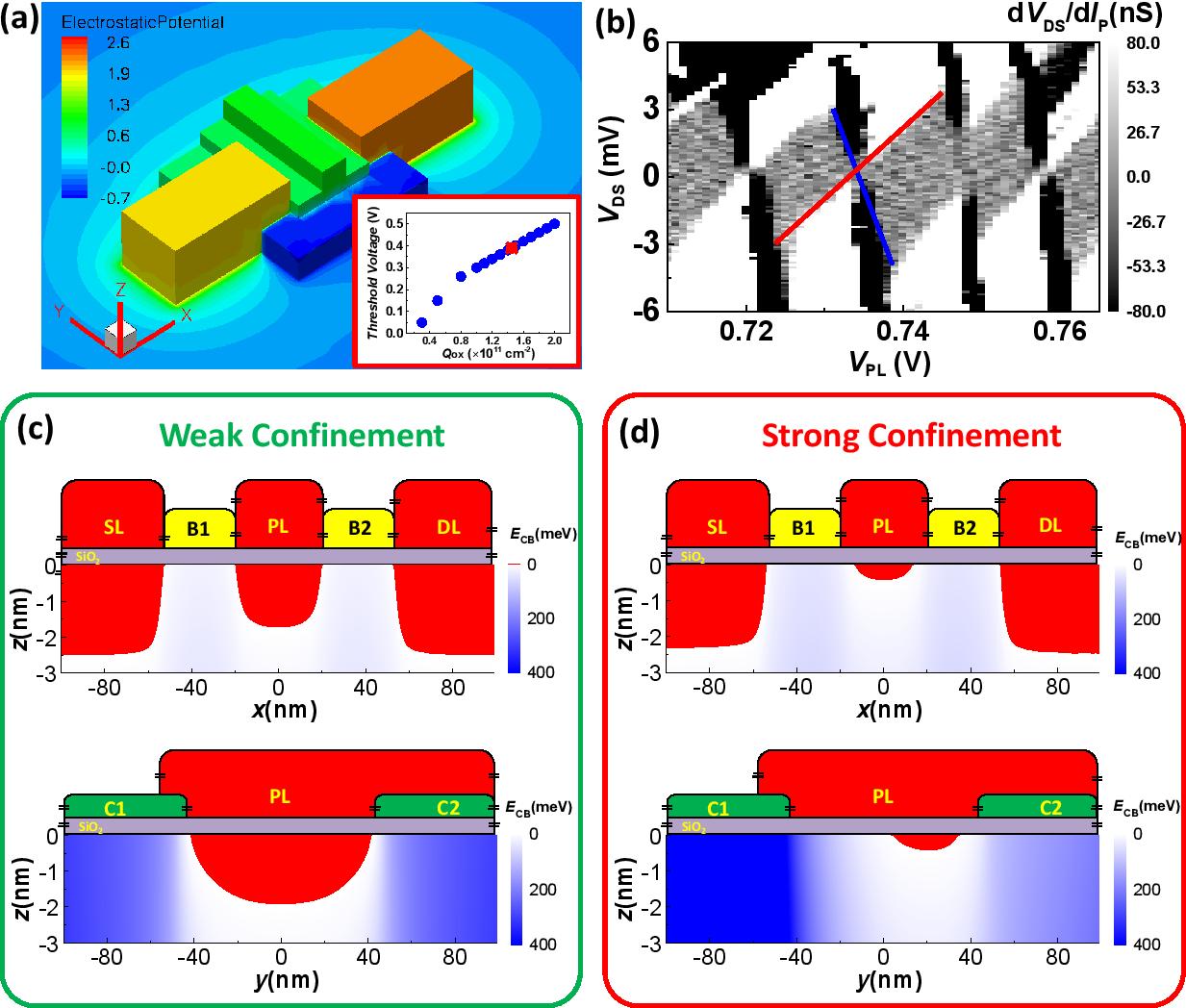}%
	\caption{(a) TCAD simulation model. Inset: The simulated threshold voltage of the device plotted as function of interface charge density $Q_\textmd{ox}$. The measured threshold voltage of $0.4$ V is achieved when $Q_\textmd{ox} = -1.45 \times 10^{11}$ cm$^{-2}$ (marked as red dot). (b) dc charge stability diagram of the device under weak electrostatic confinement. The grey diamond area is the coulomb blockaded region. $V_{\textmd{SL}} = V_{\textmd{DL}} = 1.5$ V, $V_{\textmd{B1}} = 0.8$ V, $V_{\textmd{B2}} = 1.4$ V, $V_{\textmd{C1}} = 0$ V, $V_{\textmd{C2}} = 0$ V. (c) Simulated conduction band energy ($E_\textmd{CB}$) profile under weak electrostatic confinement. We used the voltage settings from (b) for this simulation except the barrier gate voltage $V_{\textmd{B1}}$ and $V_{\textmd{B2}}$. They are adjusted to compensate the effect of metal gate induced strain at the interface. The red region in bulk silicon represents the space where $E_\textmd{CB}$ is below fermi-level. The addition energy estimated based on the simulation is $3$ meV, matching up with the value we measured in (b). Gate voltage settings for (c): $V_{\textmd{SL}} = V_{\textmd{DL}} = 1.5$ V, $V_{\textmd{B1}} = 0.34$ V, $V_{\textmd{PL}} = 0.73$ V, $V_{\textmd{B2}} = 0.34$ V, $V_{\textmd{C1}} = 0$ V, $V_{\textmd{C2}} = 0$ V, $V_{\textmd{SD}} = 0$. (d) Simulated $E_\textmd{CB}$ profile under strong electrostatic confinement (SE pumping regime). All gate voltages of the high-accuracy SE pumping experiment, presented in Fig.~\ref{fig:HighResMeas}(a), are used in this simulation except for gate B1 and B2 where the calibrated value of $V_{\textmd{B1}}$ and $V_{\textmd{B2}}$ are adopted instead. The addition energy of the QD is estimated as $17$ meV. Gate voltage settings for (d): $V_{\textmd{SL}} = 1.5$ V, $V_{\textmd{DL}} = 1.9$ V, $V_{\textmd{B1}} = V_{\textmd{B2}} = 0.34$ V, $V_{\textmd{PL}} = 0.525$ V, $V_{\textmd{C1}} = -1.04$ V, $V_{\textmd{C2}} = 0.187$ V, $V_{\textmd{SD}} = 0$ V. We do not consider the rf power applied to B1 in these simulations.\label{fig:tcad}}
\end{figure}

From the gradient of the edge of the coulomb blockade region, marked with the red solid line in Fig.~\ref{fig:tcad}(b), we acquire the ratio
$m_+ = C_\textmd{g}/(C_\Sigma-C_\textmd{s})$,  where $C_\textmd{g}$ is the capacitance between the QD and gate PL, $C_\Sigma$ is the total capacitance between the QD and surrounding environment and $C_\textmd{s}$ is the capacitance between the QD and source reservoir. Similarly, we can derive the ratio $m_- = C_\textmd{g}/C_\textmd{s}$ from the edge marked with the blue solid line in Fig.~\ref{fig:tcad}(b). Combining these two expressions gives the ratio between the capacitances $C_\Sigma$ and $C_\textmd{g}$. In our previous study \cite{Ale2014}, we have experimentally demonstrated, using a device with identical design, that the edge gradient of the coulomb blockade region is insensitive to the changes in the confinement gate voltages $V_\textmd{C1}$ and $V_\textmd{C2}$. This implies that the ratio $C_\Sigma/C_\textmd{g}$ does not change significantly when electrostatic confinement increases. We also confirm, in the measurement of another identical SE pump, that the ratio $C_\Sigma/C_\textmd{g}$ does not change with $V_\textmd{PL}$ even in the few-electron regime. We estimate the gate capacitance $C_\textmd{g}$ using the parallel plate capacitance model, $C_g=\epsilon \times A/d$, where $\epsilon$ is the dielectric constant of SiO$_2$, $d$ is the thickness of the gate oxide and $A$ is the area of the QD in the plane of the Si/SiO$_2$ interface. We acquire the parameter $A$ from the sliced 2D profile of the simulated $E_\textmd{CB}$ $0.1$ nm below the interface. Based on the dot geometry derived from TCAD simulations, the tunnelling capacitance of the entrance barrier, $C_s$, is calculated to be over an order of magnitude smaller than the dot capacitance to the PL gate and, therefore, is ignored in our estimate of the charging energy.

\section{High-Accuracy Measurement}

\subsection{On-Off Cycle Averaging Technique}
\begin{figure}
	\includegraphics[width=\linewidth]{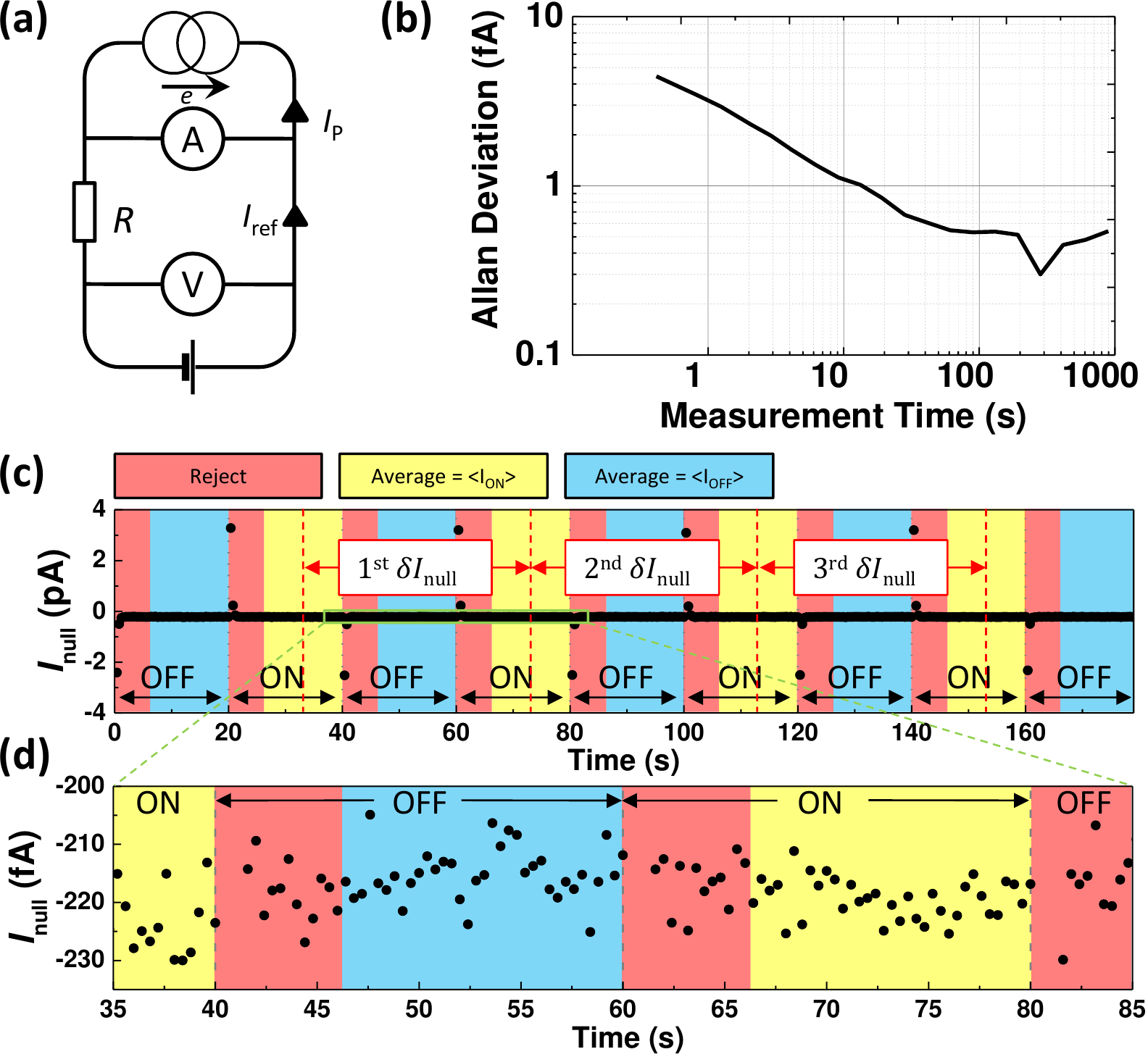}%
	\caption{(a) Circuit diagram for the high-accuracy measurement. The current, $I_{\textmd{P}}$, generated by the SE pump represented as the current source is balanced with a reference current $I_{\textmd{ref}}$, and hence only a small difference current, $I_\textmd{null}$ is measured by the ammeter. The calibrations of the 1-G$\Omega$ resistor and the voltmeter provide tracebility to primary standards. (b) Allan deviation as a function of the measurement time obtained from a $20$ minute time-series of $I_\textmd{null}$ data. (c,d): raw data for on-off cycles of $I_{\textmd{null}}$. Colored boxes indicate the ranges of data rejected or averaged to calculate the ON-OFF difference $\delta I_{\text{null}} = \langle I_{\text{ON}}\rangle - \langle I_{\text{OFF}}.\rangle$ \label{fig:HighResSetup}}
\end{figure}

Figure \ref{fig:HighResSetup}(b) shows the Allan deviation of the pump current as a function of averaging time obtained from a time series of continuously acquired data with the pump and reference currents both turned on. There is a clear transition from white (frequency-independent) noise, yielding a $1/ \sqrt{t}$ behavior, to $1/f$ noise, yielding a flat behavior, for averaging times longer than $\sim 40$~seconds. This means that further averaging of the continuously measured pump current for times longer than this will not yield any further reduction of the statistical uncertainty\cite{allan1987should}. The $1/f$ noise spectrum for long averaging times is caused by drift in offset currents and voltages present in the measurement circuit, for example the input offset bias current of the current preamplifier. As is standard practice in all types of precision electrical measurement, we reject experimental offsets by switching the current between two levels (in our case, on and off) with a cycle time within the white noise regime, and measuring the difference signal. Each ON or OFF segment consists of $50$ data points, each acquired with an integration time of $0.4$~s. Raw data from several ON-OFF measurement cycles are presented in \ref{fig:HighResSetup}(c) and \ref{fig:HighResSetup}(d). After deleting $15$ data points following each current switch to reject transient effects, the remaining data points from each OFF segment, and the two adjacent ON half-segments are averaged to yield $\langle I_{\text{OFF}}\rangle$ and $\langle I_{\text{ON}}\rangle$ respectively. The on-off difference signal $\delta I_{\text{null}} = \langle I_{\text{ON}}\rangle - \langle I_{\text{OFF}}\rangle$, and the equivalent reference current difference $\delta I_{\text{ref}}$ are used to calculate the unknown pump current.

\subsection{Uncertainty Budget of $I_{\textmd{P}}$ Measurement}
The detailed breakdown of the uncertainty budget for the measurement of $I_\textmd{P}$ is shown in Table~\ref{tab:1}. Following standard metrological practice, uncertainty terms are distinguished as either type A (statistically evaluated) or type B (evaluated by other means). The informal categorization of uncertainties as 'random' or 'systematic' correspond in most cases to type A and type B respectively. Any uncertainties less than $0.01$~ppm, for example the uncertainty in the reference frequency input to the RF source, were neglected. In the following, each component of the uncertainty budget is discussed in more detail. \vspace{5mm}

1. Voltmeter type A. The Voltmeter scale factor is calibrated before and after each measurement run directly against a Josephson voltage standard. The voltmeter is calibrated \textit{in situ}, and a low-thermal switch is used to connect its input terminals either across the voltage source, as shown in Fig.~S2(a), or across the output of the Josephson array. A typical voltmeter calibration lasts around $20$~minutes at a calibration voltage of $0.5$~V. The type A uncertainty is evaluated as the standard error of the mean of the individual on-off cycles forming the calibration. \vspace{5mm}

2. Voltmeter linearity. Because the voltmeter is calibrated at a higher voltage ($0.5$~V) than it is measuring during the experiment ($0.16$~V), some additional calibrations were performed over a range of voltages to estimate the non-linearity. The linearity correction is zero, with an uncertainty conservatively estimated as $0.03$~ppm. This uncertainty term could be eliminated by performing the calibration at the same voltage as is measured during the experiment. However, this would be at the expense of a larger statistical contribution to the calibration uncertainty. \vspace{5mm}

3. Voltmeter drift. The voltmeter was calibrated before and after each measurement run, yielding scale factors $S_{\text{V1}}$ and $S_{\text{V2}}$ respectively. The raw data was analysed using the arithmetic mean of these two scale factors to correct the voltmeter readings, and a type B uncertainty $U_{\text{drift}}$ was calculated assuming that $S_{\text{V1}}$ and $S_{\text{V2}}$ define the boundaries of a square distribution: $U_{\text{drift}}=|S_{\text{V1}}-S_{\text{V2}}|/(2 \sqrt{3})$. \vspace{5mm}

4. $1$~G$\Omega$ type B. The resistor is calibrated in a $100:1$ ratio measurement against a $10$~M$\Omega$ resistor using a cryogenic current comparator (CCC). The $10$~M$\Omega$ resistor is in turn calibrated against the quantum Hall resistance in $5$ further CCC measurement steps. The $0.1$~ppm type B uncertainty is the uncertainty in the value of the $10$~M$\Omega$ resistor in terms of the quantum Hall resistance. Uncertainty in the $100:1$ CCC current ratio contributes less than $0.01$ ppm and is not considered. \vspace{5mm}

5.  1 G$\Omega$ drift between cals. This term was evaluated in the same way as term $3$. \vspace{5mm}

6. 1 G$\Omega$, type A. Typical resistor calibrations lasted around $12-15$~hours, with the calibration composed of hundreds of cycles each lasting $\sim 100$~seconds. The standard error of the mean of the individual calibration cycles if of order $0.01$~ppm. However, as noted in the main text, the short-term stability of thick-film standard resistors is currently a subject of investigation and we do not assume \textit{a priori} that these long calibrations are sampling a stationary mean. We calculate the Allan deviation of the individual calibration cycles, and conservatively evaluate the type A uncertainty as the Allan deviation for averaging times of $\sim 2$~hours, the longest length of time for which the Allan deviation can be reliably calculated from the calibration dataset. \vspace{5mm}

7. $I_{\textmd{null}}$, type B. The $0.01 \%$ calibration uncertainty of the preamp contributes to the overall uncertainty because the preamp measures a finite current difference, $\sim 3$~fA, or $\sim 20$~ppm of the $160$~pA pumped current. The product of the uncertainty and the fractional signal is $10^{-4} \times (2\times 10^{-5})= 2\times 10^{-9}$, and we conservatively assign an uncertainty of $0.01$~ppm. \vspace{5mm}

8. $I_{\textmd{P}}$, type A. As noted in the main text, the statistical uncertainty in the pumped current, averaged over a plateau, is evaluated as the standard error of the mean of the individual data points making up the plateau. This is the largest contribution to the overall uncertainty, and is due primarily to thermal noise in the 1 G$\Omega$ reference resistor. The standard error of the mean is evaluated, following standard practice, as $\sigma / \sqrt{N}$, where the sample standard deviation $\sigma$ is defined by $\sigma^2 = \frac{1}{\sqrt{N-1}}\sum(x_{i}-\bar{x})^2$. Here $x_i$ are the $N$ individual measurements and $\bar{x}$ is the mean. For completeness, we note that the sample standard deviation slightly under-estimates the standard deviation of the underlying population. The bias in the standard deviation is roughly $4 \%$ for the case of $7$ data points, and $3 \%$ for the case of $8$ data points.

\subsection{Statistical analysis of plateau data}

As described in the main text, the plateau is defined with reference to a theoretical fit to the pump current as a function of the scanned parameter (barrier or plunger gate voltage). The plateau is defined as the range of the scanned parameter over which the fit is within $3 \times 10^{-8}$ of $ef$. Data measured with the scanned parameter within this range is considered to be 'on-plateau', and is averaged to yield a single, low-uncertainty value of the pump current. However, we cannot exclude the possibility that error processes not included in the thermal fit model influence the current in the plateau region, or that the pump current is affected by drift due to, for example $1/f$ charge noise or some other process not accounted for in the uncertainty evaluation. For this reason, we apply the same analysis to the plateau data as was used in an earlier paper \cite{giblin2017robust}. This analysis asks the question, is the distribution of $\Delta I_{\text{P}}$  data points on the plateau consistent with the hypothesis that all the data are sampling the same distribution? 

\begin{figure}
	\includegraphics[width=10cm]{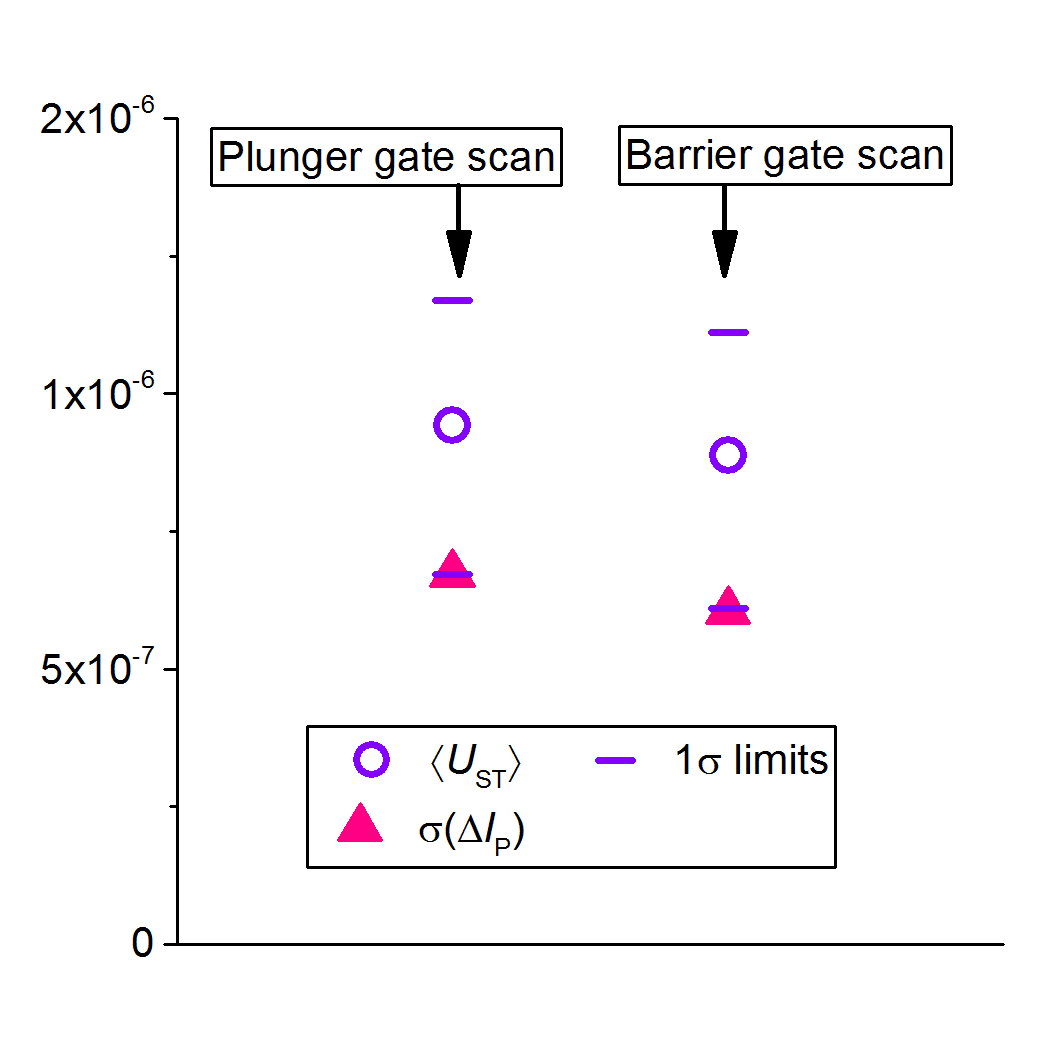}%
	\caption{Open circles: Mean statistical uncertainty  $\langle U_{\text{ST}}\rangle$ for the data points on-plateau, selected from the two measurement runs presented in figure 4 of the main text. Horizonal lines: upper and lower $1 \sigma$ limits for the distribution of the standard deviation, if $N_{\text{P}}$ points are selected from a parent distribution with standard deviation given by $\langle U_{\text{ST}}\rangle$. Filled triangles: measured standard deviation $\sigma(\Delta I_{\text{P}})$ of the data points on-plateau.\label{fig:PlateauStatistics}}
\end{figure}

Each data point in Figure 4 of the main text is averaged from $100$ on-off cycles. The statistical uncertainty (black error bars in Fig. 4 of the main text) is denoted $U_{\text{ST}}$. This uncertainty is due mainly to the thermal noise in the $1$~G$\Omega$ reference resistor, and does not change by more than $\sim 10$~$\%$ from one data point to the next. We denote the mean value of $U_{\text{ST}}$ on the plateau as $\langle U_{\text{ST}} \rangle$, and the standard deviation of the $N_{\text{P}}$ values of $\Delta I_{\text{P}}$ as $\sigma(\Delta I_{\text{P}})$. We consider each high-precision measurement as drawing $N_{\text{P}}$ samples from a parent distribution with standard deviation $\langle U_{\text{ST}} \rangle$. We calculate the probability distribution of the standard deviation of these $N_{\text{P}}$ data points using the standard result that the variance of $N$ samples is distributed according the $\chi ^2$ distribution with $N-1$ degrees of freedom. In Fig. \ref{fig:PlateauStatistics} we plot $\sigma(\Delta I_{\text{P}})$, $U_{\text{ST}}$, and the upper and lower $1 \sigma$ limits of the expected distribution of $U_{\text{ST}}$ for both the high-resolution scans illustrated in Figure 4 of the main text. For both the scans, $\sigma(\Delta I_{\text{P}})$ is within the $1 \sigma$ boundaries of the expected distribution, and we conclude that the data of Figure 4 cannot be distinguished from data randomly selected from a uniform parent distribution. It is apparent that both the scans show scatter noticeably smaller than the mean statistical uncertainty of a single data point, but there is no reason to assign significance to this observation based on just two data sets. An observation of $\sigma(\Delta I_{\text{P}})$ outside the upper $1 \sigma$ limit would be more problematic, indicating some additional structure on the plateau, or a noise process not accounted for in the uncertainty evaluation.

\bibliography{ChargePumpBibliography}